\documentclass[a4paper,USenglish,cleveref, autoref, thm-restate]{lipics-v2021}

\usepackage{todonotes}

\title{Parameterized Complexities of Dominating and Independent Set Reconfiguration\thanks{Corresponding author: Hans L. Bodlaender, h.l.bodlaender@uu.nl}}

\author{Hans L. Bodlaender}{Department of Information and Computing Sciences, Utrecht University, the Netherlands\\
Corresponding author}{h.l.bodlaender@uu.nl}{https://orcid.org/0000-0002-9297-3330}{}

\author{Carla Groenland}{Faculty of Electrical Engineering, Mathematics and Computer Science. Technical University Delft, the Netherlands}{c.e.groenland@uu.nl}{
http://orcid.org/0000-0002-9878-8750}{
Supported by the European Union’s Horizon 2020 research and innovation programme under the ERC grant CRACKNP (number 853234) and the Marie Skłodowska-Curie grant GRAPHCOSY (number 101063180). The research was done when Carla Groenland was associated with Utrecht University.}
\author{C\'eline M. F. Swennenhuis}{Department of Mathematics and Computer Science, Eindhoven University of Technology, The Netherlands}{c.m.f.swennenhuis@tue.nl}{
https://orcid.org/0000-0001-9654-8094}{
Supported by the Netherlands Organization for Scientific Research under project no. 613.009.031b.}

\authorrunning{H.\,L. Bodlaender, C. Groenland and C.\,M.\,F. Swennenhuis} 

\Copyright{Hans L. Bodlaender, Carla Groenland, C\'eline M. F. Swennenhuis} 


\relatedversion{This paper combines the results reported in \cite{BodlaenderGS21} at IPEC 2021 and some of the results reported in \cite{BodlaenderGNS21} at FOCS 2021.}


\category{} 

\nolinenumbers 

\hideLIPIcs  

\usepackage[utf8]{inputenc}
\usepackage{amsmath,amssymb}
\usepackage{amsthm}
\usepackage{todonotes,url}
\usepackage{color}
\usepackage{romannum}
\newtheorem{problem}[theorem]{Problem}

\begin{document}
\pagenumbering{arabic}
\maketitle
\newcommand{\N}{{\bf N}} 
\newcommand{\Z}{{\bf Z}} 
\newcommand{\longchain}{{\sc Long Partitioned Positive Chain Satisfiability}}

\newcommand{\NP}{\mathrm{NP}}
\newcommand{\XNLP}{\mathrm{XNLP}}
\newcommand{\XNL}{\mathrm{XNL}}
\newcommand{\XL}{\mathrm{XL}}
\def\poly{\operatorname{poly}}

\begin{abstract}
    We settle the parameterized complexities of several variants of independent set reconfiguration and dominating set reconfiguration, parameterized by the number of tokens. We show that both problems are XL-complete when there is no limit on the number of moves, XNL-complete when a maximum length $\ell$  for the sequence is given in binary in the input, and XNLP-complete when $\ell$ is given in unary. The problems were known to be
   $\mathrm{W}[1]$- and $\mathrm{W}[2]$-hard respectively when $\ell$ is also a parameter. We complete the picture by showing membership in those classes. 
    
    Moreover, we show that for all the variants that we consider, token sliding and token jumping are equivalent under pl-reductions. We introduce partitioned variants of token jumping and token sliding, and give pl-reductions between the four variants that have precise control over the number of tokens and the length of the reconfiguration sequence.
\end{abstract}

\section{Introduction}
In this paper, we study the parameterized complexity of reconfiguration of independent sets, and of
dominating sets, with the sizes of the sets as parameter. Interestingly, the complexity varies depending on the assumptions on the length of the reconfiguration sequence, which can be unbounded, given in binary, given in unary, or given as second parameter. One can study the reconfiguration problems for different reconfiguration rules; we will show
equivalence regarding the complexity for several reconfiguration rules.

\paragraph*{Independent Set Reconfiguration}

In the {\sc Independent Set Reconfiguration} problem, we are given a graph and two independent sets $A$ and $B$, and wish to decide we can `reconfigure' $A$ to $B$ via a `valid' sequence of independent sets $A,I_1,\dots,I_{\ell-1},B$.
Suppose that we represent the current independent set by placing a token on each vertex. We can move between two independent sets by moving a single token. We consider two well-studied rules for deciding how we can move the tokens.
\begin{itemize}
    \item Token jumping (TJ): we can `jump' a single token to any vertex that does not yet contain a token. 
    \item Token sliding (TS): we can `slide' a single token to an adjacent vertex that does not yet contain a token.
\end{itemize}
\textsc{Independent Set Reconfiguration} is PSPACE-complete for both rules \cite{IDHP11,HD05,Kaminski12PSPACE}, but their complexities may be different when restricting to specific graph classes. For example, \textsc{Independent Set Reconfiguration} is NP-complete on bipartite graphs under the token jumping rule, but remains PSPACE-complete under the token sliding rule \cite{Lokshtanov18bipartite}.  $W[1]$-hardness for TJ when parameterized by the number of tokens was shown in \cite{IKOSUY20}.

There is a third rule which has been widely studied, called the token addition-removal rule, but this rule is equivalent to the token jumping rule for our purposes (see e.g. \cite[Theorem 1]{Kaminski12PSPACE}). As further explained later, we show that the token jumping and the token sliding rule are also equivalent in some sense (which is much weaker but does allow us to control all the parameters that we care about). We will therefore not explicitly mention the specific rule under consideration below.

Throughout this paper, our reconfiguration problems are parameterized by the number of tokens (the size of the independent set). \textsc{Independent Set Reconfiguration} is $\mathrm{W}[1]$-hard \cite{Ito14}, but the problem is not known to be in $\mathrm{W}[1]$. We show that in fact it is complete for the class XL, consisting of the parameterized problems that can be solved by a 
deterministic algorithm that uses $f(k) \log n$ space, where $k$ is the parameter, $n$ the input size and $f$ any computable function. 
\begin{theorem}
\label{thm:XLmain}
\textsc{Independent Set Reconfiguration} is $\XL$-complete. 
\end{theorem}

In the \textsc{Timed Independent Set Reconfiguration}, we are given an integer $\ell$ in \emph{unary} and two independent sets $A$ and $B$ in a graph $G$, and need to decide whether there is a reconfiguration sequence from $A$ to $B$ of length at most $\ell$. We again parameterize it by the number of tokens. We show that this problem is complete for the class XNLP. 
The class XNLP (also denoted $N[f \poly, f \log]$ by
Elberfeld et al.~\cite{ElberfeldST15}) is the class of parameterized problems that can be solved with a non-deterministic algorithm with simultaneously, the running
time bounded by $f(k)n^c$ 
and the space usage bounded by $f(k)\log n$, with $k$ the parameter, $n$ the
input size, $c$ a constant, and $f$ a computable function. 

\begin{theorem}
\textsc{Timed Independent Set Reconfiguration} is $\XNLP$-complete.
\end{theorem}

XNLP is a natural subclass of the class XNL, which consists of the parameterized problems that can be solved by a 
nondeterministic algorithm that uses $f(k) \log n$ space. Amongst others, XNL was studied by Chen et al.~\cite{ChenF03}. 
The classes XL, XNL, XSL, XP can be seen as the parameterized counterparts of L, NL, SL, P respectively.  Although no explicit time bound is given, we can freely add a time bound of $2^{f(k)\log n}$, and thus XNL is a subset of XP.  
We remark that XL$=$XSL (in the same way as L$=$XL; see \cite{wehar2017complexity} and Appendix~\ref{app:STM}), XL$\subseteq $XNL and XNLP$\subseteq $XNL. 

In \textsc{Binary Timed Independent Set Reconfiguration}, the bound $\ell$ on the length of the sequence is given in binary\footnote{Giving $\ell$ in binary implies that it contributes $\log_2(\ell)$ to the size of an instance of \textsc{Binary Timed Independent Set Reconfiguration}.  }. Interestingly, this slight adjustment to \textsc{Timed Independent Set Reconfiguration} is complete for XNL instead. 
\begin{theorem}
\label{thm:xnl_main}
\textsc{Binary Timed Independent Set Reconfiguration} is $\XNL$-complete. 
\end{theorem}

Finally, we consider what happens when we consider $\ell$ to be a parameter instead. Consider \textsc{Timed Independent Set Reconfiguration} (or equivalently \textsc{Binary Timed Independent Set Reconfiguration}) is parameterized
by the size of the independent set and the length of the sequence\footnote{We can also consider it to be parameterized by the sum of the two parameters.}. Mouawad et al.~\cite{MNRSS17} showed that this problem
is $\mathrm{W}[1]$-hard\footnote{Mouawad et al. \cite{MNRSS17} only studied the token jumping variant, but Theorem \ref{thm:equivalences} implies the hardness also holds for token sliding.}. 
We show that in this case, $\mathrm{W}[1]$ is the `correct class'.
\begin{theorem}
\label{thm:W1main}
\textsc{Timed Independent Set Reconfiguration} is in $\mathrm{W}[1]$ when parameterized by the size of the independent set and the length of the sequence.
\end{theorem}

\begin{table}
    \centering
    \begin{tabular}{l|l|l|l}
 Sequence length $\ell$ & Independent Set & Dominating Set  & Sources \\
    \hline
    not given & XL-complete  & XL-complete & Section~\ref{section:is-xl}, \ref{section:ds-xl} \\
    parameter     & $\mathrm{W}[1]$-complete  & $\mathrm{W}[2]$-complete  & Sections~\ref{section:is-w}, \ref{section:ds-w} and \cite{MNRSS17}\\
    unary input & XNLP-complete  & XNLP-complete & Section~\ref{section:is-xnlp}, \ref{section:ds-xnlp} \\
    binary input & XNL-complete  & XNL-complete & Section~\ref{section:is-xnl}, \ref{section:ds-xnl} \\
    \end{tabular}
    \caption{The table shows the parameterized complexities of the independent set and dominating set reconfiguration problems, parameterized by the number of tokens, depending on the treatment of the bound $\ell$ on the length of the reconfiguration sequence.}
    \label{tab:picture}
\end{table}

\begin{table}
    \centering
    \begin{tabular}{c|c|c|c}
    Complexity class     & Non-deterministic? & Time constraints & Space constraints  \\
    \hline
    XL & No & None & $f(k)\log n$\\
    XNL & Yes & None &$f(k)\log n$\\
    XNLP & Yes & $f(k)n^{O(1)}$&$f(k)\log n$\\
    \end{tabular}
    \caption{Given a parameterized problem, we write $k$ for the parameter and $n$ for the input size. An overview of the classes XL, XNL and XNLP is given in the table: each is the class of parameterized problems defined via some time/space constraints of a (non-deterministic or deterministic) Turing machine.}
    \label{tab:my_label}
\end{table}

\paragraph*{Dominating set reconfiguration}
The dominating set reconfiguration problem is similar to the independent set reconfiguration problem, but in this case all sets in the sequence must form a dominating set in the graph. This again gives a PSPACE-complete problem \cite{Ito12ea},
even for simple graph classes such as planar graphs and classes of bounded bandwidth \cite{HIMNOST16}, see also \cite{BDO19}.
We define the parameterized problems \textsc{Dominating Set Reconfiguration}, \textsc{Timed Dominating Set Reconfiguration}  and \textsc{Binary Timed Dominating Set Reconfiguration} similarly as their independent set counterparts, again parameterized by the number of tokens.
Since \textsc{Dominating Set} is $\mathrm{W}[2]$-complete and \textsc{Independent Set} is $\mathrm{W}[1]$-complete (parameterized by `the number of tokens'), it may be expected that the reconfiguration variants also do not have the same parameterized complexity. Indeed,  it is known that
\textsc{Timed Dominating Set Reconfiguration} is $\mathrm{W}[2]$-hard when it is moreover parameterized by the length of the sequence \cite{MNRSS17}. We complement this result by showing that is belongs to $W[2]$ (Section~\ref{section:ds-w}), and show that the picture is otherwise the same as for \textsc{Independent Set}: \textsc{Dominating Set Reconfiguration} is $\XL$-complete (Section~\ref{section:ds-xl}),
\textsc{Timed Independent Set Reconfiguration} is $\XNLP$-complete (Section~\ref{section:ds-xnlp}), and
\textsc{Binary Timed Dominating Set Reconfiguration} is $\XNL$-complete (Section~\ref{section:ds-xnl}).


A summary of our results can be found in Table \ref{tab:picture}.

Many other types of reconfiguration problems have been studied as well, and we refer the reader to the surveys by Van den Heuvel \cite{vandenheuvel} and Nishimura \cite{nishimura} for further background.

\paragraph*{Equivalences between token jumping and token sliding}
In Section~\ref{section:equivalences}, 
we introduce partitioned variants of token sliding and token jumping in which the tokens need to stay within specified token sets.
Both for the partitioned and the regular variant, and for the variant with the token jumping rule and the token sliding rule,
we prove equivalence under pl-reductions and fpt-reductions for each of the following problems:
\textsc{Independent Set Reconfiguration}, \textsc{Timed Independent Set Reconfiguration}, \textsc{Binary Independent Set Reconfiguration} and \textsc{Timed Independent Set Reconfiguration} when moreover parameterized by the length of the sequence. The same holds for the dominating set variants.



\section{Preliminaries}
\label{sec:preliminaries}

We write $\N$ for the set of integers $0, 1, 2,\dots$ and write $[a,b]$ for the set of integers $x$ with $a\leq x \leq b$. All $\log$s in this paper are base 2.

\subsection{Parameterized reductions}
A {\em parameterized reduction} from a parameterized problem $Q_1 \subseteq \Sigma_1^\ast \times \N$ to a parameterized problem $Q_2 \subseteq \Sigma_2^\ast \times \N$ is a function
$f : \Sigma_1^\ast \times \N \rightarrow \Sigma_2^\ast \times \N$, such that the following holds.
\begin{enumerate}
    \item For all $(x,k) \in \Sigma_1^{\ast} \times \N$, $(x,k)\in Q_1$ if and only if $f((x,k)) \in Q_2$.
    \item There is a computable function $g$, such that for all $(x,k) \in \Sigma_1^\ast \times \N$, if $f((x,k)) = (y,k')$, then $k' \leq g(k)$.
\end{enumerate}
A {\em parameterized logspace reduction} or {\em pl-reduction} is a parameterized reduction
for which there is an algorithm that computes $f((x,k))$ in space $\mathcal{O}(g(k) + \log n)$, with $g$ a computable
function and $n=|x|$ the number of bits to denote $x$.

\subsection{Turing machines and complexity classes}
We assume the reader to be familiar with the basics of the notion of Turing Machines and Non-deterministic Turing Machines.
We discuss a few relevant aspects, and notation, that will be used in our proofs.

\subsubsection{Nondeterministic Turing Machines and XNL}
Recall that a Nondeterministic Turing Machine (NTM) with one work tape is a 5-tuple $(\mathcal{S},\Sigma,\mathcal{T},s_{\sf start},\mathcal{A})$, where
$\mathcal{S}$ is a finite set of \emph{states}, $\Sigma$ is the \emph{alphabet}, $\mathcal{T}$ is the set of \emph{transitions},
    $s_{\sf start}$ is the \emph{start state} and
    $\mathcal{A}$ is the set of \emph{accepting states}.
    
Normally, Turing Machines are defined to have an input $\alpha\in\Sigma^*$ on the input tape. The input tape is (in contrast with the work tape) immutable. A transition $T\in \mathcal{T}$ is then a tuple of the form $(p,\Delta_{\sf work}, \Delta_{\sf inp} ,q)$, where $\Delta_{\sf work}$ and $\Delta_{\sf inp}$ are both tape triples describing how the transition affects the work tape and input tape respectively. As the input tape is immutable, any $\Delta_{\sf inp}$ will be of the form $(e,\delta,e)$ with $\delta\in\{-1,0,1\}$. We say that $\alpha\in\Sigma^*$ is \emph{accepted} by an NTM $\mathcal{M}$ if the computation of $\mathcal{M}$ with $\alpha$ on the input tape ends in an accepting state. 
The class XNL consists of the parameterized problems accepted by a NTM with a work tape of size
$f(k)\log n$, with $n$ the input size, $k$ the parameter, and $f$ a computable function.
XL is defined in the same way, using now a deterministic TM. XNLP is the class of
parameterized problems accepted by a NTM in $f(k)n^{O(1)}$ time with a work tape of size
$f(k)\log n$, with $n$, $k$, and $f$ as above. 

Following Elberfeld et al.~\cite{ElberfeldST15}, we can also use a work tape with $k$ cells and a larger 
alphabet. The following problem is the starting point for our reductions for XNL-hardness.

\begin{verse}
    {\sc Input Accepting Log-Space Nondeterministic Turing Machine}\\
    {\bf Given:} An NTM $\mathcal{M} = (\mathcal{S},\Sigma,\mathcal{T},s_{\sf start},\mathcal{A})$ with $\Sigma = [1,n]$, a work tape with $k$ cells and input $\alpha \in \Sigma^*$. \\
    {\bf Parameter:} $k$.\\
    {\bf Question:} Does $\mathcal{M}$ accept $\alpha$? 
\end{verse}

The {\sc Input Accepting Log-Space Deterministic Turing Machine} problem is defined similar, except
that we use a deterministic TM.

\begin{theorem}[Elberfeld et al.~\cite{ElberfeldST15}]
{\sc Input Accepting Log-Space Nondeterministic Turing Machine} is XNL-complete, and
{\sc Input Accepting Log-Space Deterministic Turing Machine} is XL-complete.
\label{theorem:xnlandxl}
\end{theorem}

\paragraph*{Symmetric Turing Machine}
A Symmetric Turing Machine (STM) is a Nondeterministic Turing Machine (NTM), where the transitions are symmetric. That means that for any transition, we can also take its inverse back. More formally, a Symmetric Turing Machine with one work tape is a 5-tuple $(\mathcal{S},\Sigma,\mathcal{T},s_{\sf start},\mathcal{A})$, where
$\mathcal{S}$ is a finite set of \emph{states}, $\Sigma$ is the \emph{alphabet}, $\mathcal{T}$ is the set of \emph{transitions},
    $s_{\sf start}$ is the \emph{start state} and
    $\mathcal{A}$ is the set of \emph{accepting states}.
A transition $\tau \in \mathcal{T}$ is a tuple of the form $(p,\Delta,q)$ describing a transition the STM may take, where $p,q \in \mathcal{S}$ are states and $\Delta$ is a \emph{tape triple}. A tape triple is equal to either $(ab,\delta,cd)$, where $a,b,c,d\in \Sigma$ and $\delta \in \{-1,1\}$, or $(a,0,b)$, where $a,b\in \Sigma$. 
For example, the transition $(p,(ab,1,cd),q)$ describes that if the STM is in state $p$, reads $a$ and $b$ on the current work tape cell and the cell directly right of it, then it can replace $a$ with $c$, $b$ with $d$, moving the head to the right and going to state $q$. 

Let $\Delta = (ab,\delta,cd)$ be a state triple, then its inverse is defined as $\Delta^{-1} = (cd,-\delta,ab)$. The inverse of $\Delta = (a,0,b)$ is defined as $\Delta^{-1} = (b,0,a)$.
By definition of the Symmetric Turing Machine, for any $\tau\in \mathcal{T}$, there is an \emph{inverse} transition $\tau^{-1}\in \mathcal{T}$, i.e. if $\tau =(p,\Delta,q)\in \mathcal{T}$, then $\tau^{-1} =(q,\Delta^{-1},p)\in \mathcal{T}$.


We can define XSL in the same way as XNL or XL, using now a STM. Similar as SL$=$L, we have that
XSL$=$XL; see the discussion in Appendix~\ref{app:STM}.

We say that STM $\mathcal{M}$ \emph{accepts} if there is a computation of $\mathcal{M}$ that ends in an accepting state. We remark that the Turing Machines in this paper do not have an input tape, as it is hidden in the states. 
For a more formal definition of Symmetric Turing Machines we would like to refer to the definition from Louis and Papadimitriou in \cite{lewis1982symmetric}. 

Note that we may assume that there is only one accepting state $s_{\sf acc} \in \mathcal{A}$, by creating this new state $s_{\sf acc}$ and adding a transition to $s_{\sf acc}$ from any original accepting state. We may also assume all transitions to move the tape head to the left or right. This can be accomplished by replacing each transition $\tau=(p,(a,0,b),q)$ with $2|\Sigma|$ transitions as follows. For all $\sigma\in \Sigma$, we create a new state $s_\sigma$ and two new transitions $\tau^1_\sigma = (p,(a\sigma,1,b\sigma),s_{\sigma})$ and $\tau^2_\sigma = (s_{\sigma},(b\sigma,-1,b\sigma),q)$. 

The following problem will be used in the reductions of Section~\ref{sec:XLIS}. 

\begin{verse}
    {\sc Accepting Log-Space Symmetric Turing Machine}\\
    {\bf Given:} A STM $\mathcal{M} = (\mathcal{S},\Sigma,\mathcal{T},s_{\sf start},\mathcal{A})$ with $\Sigma = [1,n]$ and a work tape with $k$ cells.\\
    {\bf Parameter:} $k$.\\
    {\bf Question:} Does $\mathcal{M}$ accept? 
\end{verse}

We define {\sc Accepting Log-Space Nondeterministic  Turing Machine} to be the Nondeterministic Turing Machine analogue of {\sc Accepting Log-Space Symmetric Turing Machine}.

\begin{theorem}
\label{theorem:xsl=xl}
{\sc Accepting Log-Space Symmetric Turing Machine} is $\XL$-complete.
\end{theorem}
We include a proof of Theorem \ref{theorem:xsl=xl} in Appendix \ref{app:STM} for completeness.

In our reductions we use the notion of a \emph{configuration}, describing exactly in what state an NTM (and therefore an STM) and its tape are. 
\begin{definition} \label{def:configuration} Let $\mathcal{M}=(\mathcal{S},\Sigma,\mathcal{T},s_{\sf start},\mathcal{A})$ be an NTM with $\Sigma=[1,n]$ and $k$ cells on the work tape and let $\alpha \in \Sigma^*$ be the input. A \emph{configuration} of $\mathcal{M}$ is a $k+2$ tuple $(p,i,\sigma_1,\dots,\sigma_k)$ where $p \in \mathcal{S}$, $i \in [1,k]$ and $\sigma_1,\dots,\sigma_k\in \Sigma$, describing the state, head position and content of the work tape of $\mathcal{M}$ respectively.
\end{definition}

\section{(Partitioned) Token Jumping and (Partitioned) Token Sliding Equivalences}
\label{section:equivalences}
Consider the following four rules for movement of tokens:
\begin{itemize}
    \item Token Sliding (TS): we can `slide' a token to an empty\footnote{We say a vertex is empty if it has no token on it (i.e. is not part of the independent set).} adjacent vertex.
    \item Partitioned Token Sliding: a partition of the vertices has been given, and we can only `slide' a token to an adjacent vertex within the same token set.
    \item Token Jumping (TJ): we can `jump' a token to an empty vertex.
    \item Partitioned Token Jumping: a partition of the vertices has been given, and we can only `jump' a token to an empty vertex within the same token set.
\end{itemize}

In this section, we show that these four problems are equivalent, in the sense as expressed by the following theorem, which summarizes the findings of this section.

\begin{theorem}
\label{thm:equivalences}
For the following  parameterized problems, 
their variant with the token jumping rule is equivalent  under pl-reductions and fpt-reductions to their variant with the token sliding rule: \textsc{Independent Set Reconfiguration}, \textsc{Timed Independent Set Reconfiguration}, \textsc{Binary Independent Set Reconfiguration} and \textsc{Timed Independent Set Reconfiguration} when moreover parameterized by the length of the sequence. The same holds for the dominating set variants.
\end{theorem}

An overview of the reductions is given in Figure~\ref{fig:TJ/TSreduction}. 
\begin{figure}[h]
\centering
\begin{tikzpicture}[scale=0.90,transform shape]
    \node[draw, rectangle, thick] (PS) at (4,1) {Partitioned Token Sliding};
    \node[draw, rectangle, thick] (PJ) at (4,-1) {Partitioned Token Jumping};
    \node[draw, rectangle, thick] (S) at (-3,1) {Token Sliding};
    \node[draw, rectangle, thick] (J) at (-3,-1) {Token Jumping};
    
    \draw[thick,->,shorten >=5pt] (PS) to [bend left=0] (PJ);
    \node at (5.5,0.25) {Lemma \ref{thm:varIS3} (IS)};
    \node at (5.5,-0.25) {Lemma \ref{thm:varDS3} (DS)};
    
    \draw[thick,->,shorten >=5pt] (PJ) to [bend left=0] (J);
    \node at (0,-1.5) {Lemma \ref{thm:varIS4} (IS)};
    \node at (0,-2) {Lemma \ref{thm:varDS4} (DS)};
    
    \draw[thick,->,shorten >=5pt] (J) to [bend left=0] (S);
    \node at (-4.5,.25) {Lemma \ref{thm:varIS1} (IS)};
    \node at (-4.5,-.25) {Lemma \ref{thm:varDS1} (DS)};
    
    \draw[thick,->,shorten >=5pt] (S) to [bend left=0] (PS);
    \node at (0,2) {Lemma \ref{thm:varIS2} (IS)};
    \node at (0,1.5) {Lemma \ref{thm:varDS2} (DS)};
    
\end{tikzpicture}
\caption{Overview of the reductions that are used to prove Theorem \ref{thm:equivalences}.}
\label{fig:TJ/TSreduction}
\end{figure}

\subsection{Equivalences for Independent Set}
\begin{lemma}
\label{thm:varIS1}\label{lemma:firstIS}
There exists a pl-reduction $f$ from {\sc TJ-Independent Set Reconfiguration} to {\sc TS-Independent Set Reconfiguration}, such that for any instance $\mathcal{A}$ of {\sc TJ-Independent Set Reconfiguration} with $k$ tokens,  $\mathcal{A}$ admits a reconfiguration sequence of length $\ell$ if and only if $f(\mathcal{A})$ admits a reconfiguration sequence of length $\ell+k+1$.
\end{lemma}
\begin{proof}
Let $\mathcal{A}=(G,I_{\sf init},I_{\sf fin},k)$ be an instance of {\sc TJ-Independent Set Reconfiguration} with $G=(V,E)$ and $k=|I_{\sf init}|$ tokens. We create an instance $f(\mathcal{A})$ of {\sc TS-Independent Set Reconfiguration} with $k$ tokens. 

For each token $i\in[1,k]$ of $\mathcal{A}$, we first create the following gadget that models on which vertex of $G$ the token is. Let $V^i=\{v^i:v\in V\}$ be a copy of $V$ and let it induce an $n$-vertex clique.
At most one vertex of $V^i$ can hence be in an independent set. Since there are $k$ tokens and $k$ token gadgets, the pigeonhole principle implies that exactly one token must be in each gadget $V^i$. If a token is placed on $v^i$, this corresponds to placing the  $i$th token on vertex $v$. 

We add an edge $v^iw^j$ for all $i,j\in[1,k]$ with $i\neq j$, when $vw\in E$ or $v=w$. This results in a graph $G'$. The edges $v^iv^j$  ensure that no two gadgets `select' the same vertex and the other edges will ensure that each independent set in $G'$ corresponds to an independent set in $G$.

There is still an issue now that for every independent set in $G$, there are $k!$ independent sets in $G'$ that correspond to it, which is an issue when defining the initial and final independent sets.\footnote{We could have solved this as well by considering $k!$-to-$1$ reductions instead. In fact, the `ordered TJ-reconfiguration graph' of $G$ is isomorphic to the `TS-reconfiguration graph' of $G'$.}
To handle this, we add $k$ more vertices $f^1,\dots,f^k$ with edges $f^i v^i$ for all $v\in V$  and $i\in [1,k]$. Furthermore, we add two vertices $z_{\sf{init}}$ and $z_{\sf fin}$,  with edges $f^iz_{\sf init}$ for all $i\in [1,k]$ and $v^iz_{\sf fin}$ for all $v \in V\setminus I_{\sf init}$. Let $G''$ be the resulting graph.

We set $I_{\sf fin}'=\{f^1,\dots,f^k,z_{\sf fin}\}$ and $I_{\sf init}'=\{(s_i)^i:i\in [1,k]\}\cup \{z_{\sf init}\}$ for $s_1,\dots,s_k$ an arbitrary order on the $k$ vertices in $I_{\sf init}$. We define $f(\mathcal{A})=(G'',I_{\sf init}',I_{\sf fin}',k+1)$. 

We omit the details why this construction works, and only give an informal explanation. A TJ-reconfiguration sequence $I_{\sf init}= I_0,I_1,\dots,I_\ell=I_{\sf fin}$ in $\mathcal{A}$ can be converted to a TS-reconfiguration sequence $I_{\sf init}'=I_0',I_1',\dots,I_\ell'$ by numbering the tokens in $I_{\sf init}$ the way we did for $I_{\sf init}'$ and `tracking' their movements to the sequence, mimicking the movements within $G''$. Once we reach some set $I_\ell'$ that `corresponds' to $I_{\sf fin}$, we can move the token from $z_{\sf init}$ to $z_{\sf fin}$ and then all the other tokens to $f^i$ for all $i\in [1,k]$ by $k$ further moves. This is the only way to arrive at $I'_{\sf fin}$: $G''$ has $k+1$ disjoint cliques ($V^1\cup \{f^1\},\dots,V^k \cup \{f^k\}$ and $\{z_{\sf init}, z_{\sf fin}\}$) and there are $k+1$ tokens, so exactly one token must be in each clique at any moment. In particular, the tokens are forced to stay in their respective cliques. The token starting at $z_{\sf init}$ must move to $z_{\sf fin}$. As long as the token is on $z_{\sf init}$, the $f^i$ vertices cannot receive a token. The token on $z_{\sf init}$ can only move to $z_{\sf fin}$ if the tokens in the other cliques correspond to the (original) final independent set $I_{\sf fin}$. Once $z_{\sf fin}$ has a token, the other tokens can move to $f^i$ for all $i\in [1,k]$. This implies that there is a sequence in $\mathcal{A}$ of length $\ell$ if and only if there is one in $f(\mathcal{A})$ of length $\ell+k+1$.
\end{proof}

\begin{lemma}\label{thm:varIS2}
There exists a pl-reduction $f$ from {\sc TS-Independent Set Reconfiguration} to {\sc Partitioned TS-Independent Set Reconfiguration}, such that for any instance $\mathcal{A}$ of {\sc TS-Independent Set Reconfiguration} with $k$ tokens,  $\mathcal{A}$ admits a reconfiguration sequence of length $\ell$ if and only if $f(\mathcal{A})$ admits a reconfiguration sequence of length $\ell+k+1$.
\end{lemma}
\begin{proof}
Let $\mathcal{A}=(G,I_{\sf init},I_{\sf fin},k)$ be an instance of {\sc TS-Independent Set Reconfiguration} with $G=(V,E)$. 
We create $k$ sets $P^1,\dots,P^k$, where each $P^i$ is a copy of $V$. We write $v^i\in P^i$ for the $i$th copy of $v\in V$. Each $P^i$ forms a token set, meaning that each independent set must contain exactly one vertex from $P^i$. This chosen vertex in $P^i$ models the choice of the $i$th token in $\mathcal{A}$. 

We add the edges $v^iw^j$ for all $i,j\in [1,k]$ with $vw\in E$ or $v=w$. This defines a graph $G'$. The proof now continues as in the proof of Lemma \ref{thm:varIS1}: there is again a $k!$-to-one correspondence between independent sets $I'$ of $G'$ and independent sets $I$ of $G$ (for $I'$ of size $k$ in $G'$, we consider the set of $v\in V$ for which $v^i\in I'$ for some $i\in [1,k]$). We obtain $G''$ from $G'$ by adding new vertices $f^1,\dots,f^k, z_{\sf init}$ and $z_{\sf fin}$ and the same edges as in the proof of Lemma \ref{thm:varIS1}, and add $f^i$ to the $i$th token set. Again, an extra token set with $\{z_{\sf init},z_{\sf fin}\}$ is created. The remainder of the analysis is analogous.
\end{proof}

\begin{lemma}\label{thm:varIS3}
There exists a pl-reduction $f$ from {\sc Partitioned TS-Independent Set Reconfiguration} to {\sc Partitioned TJ-Independent Set Reconfiguration}, such that for any instance $\mathcal{A}$ of {\sc Partitioned TS-Independent Set Reconfiguration},  $\mathcal{A}$ admits a reconfiguration sequence of length $\ell$ if and only if $f(\mathcal{A})$ admits a reconfiguration sequence of length $3\ell$.
\end{lemma}
\begin{proof}
Let $\mathcal{A}=(G,I_{\sf init},I_{\sf fin},k)$ be an instance of {\sc Partitioned TS-Independent Set Reconfiguration} with $G=(V,E)$ and $P_1,\dots,P_k$ the tokens sets.

Let $i\in [1,k]$. We create two vertex sets $A_i$ and $B_i$ that contain a copy $v^a$ and $v^b$ respectively of each vertex $v \in P_i$. We add edges $v^aw^b$ for all $v,w \in P_i$ with $v \neq w$, as well as edges $v^aw^a$ for $vw\in E$.
Finally, for each $vw\in E(G[P_i])$, we add a vertex $\delta_{vw}$ that we connect to  $v^b$ and $w^b$, and we connect all $\delta_{vw}$ vertices to each other. Let $G'$ be the resulting graph. We model the $i$th token being on vertex $v$ in $\mathcal{A}$ by a token on $v^a$.

In any independent set $I'$ of $G'$, either $v^a$ and $v^b$ are in $I'$ for some $v\in P_i$, or $v^a$ and $\delta_{vw}$ are in $I'$ for some $v\in P_i$ and $vw\in E(G[P_i])$. 
Moreover, $I'$ can contain at most one $\delta_{vw}$ vertex. To any independent set $I$ in $G$, we correspond the independent set $g(I)$ of size $2k$ in $G'$ given by $\{v^a,v^b:v\in I\}$.
We create an instance $f(\mathcal{A})=(G',g(I_{\sf init}), g(I_{\sf fin}),2k)$. The token sets are given by the partition $P_1',\dots,P_{2k}'$ with for $i\in [1,k]$, $P_i'=A_i$ and 
$$
P_{k+i}'=B_i\cup \{\delta_{vw}:vw\in E(G[P_i])\}.
$$
A slide from $v$ to $w$ (say from $I_1$ to $I_2$) over an edge $vw\in E(G[P_i])$ corresponds to the sequence of jumps $v^b \to \delta_{vw}$, $v^a \to w^a$, $\delta_{vw} \to w^b$ (say from $g(I_1)$ to $g(I_2)$). 
For each reconfiguration sequence $I_0,\dots,I_\ell$ in $\mathcal{A}$ of length $\ell$, there is a reconfiguration sequence of length $3\ell$ in $f(\mathcal{A})$ of the form
$g(I_0),I'_{01},I'_{02},g(I_1),I'_{11},\dots, g(I_{\ell})$. Conversely, if there is no reconfiguration sequence between $I_{\sf init}$ and $I_{\sf fin}$ of length at most $\ell$, then any sequence in $f(\mathcal{A})$ from $g(I_{\sf init})$ to $g(I_{\sf fin})$ must have a subsequence of length at least $\ell+1$ consisting of distinct independent sets $g(I_0),\dots,g(I_\ell)$, and between any two must be at least two more independent sets not of the form $g(I)$. Hence there is also no reconfiguration sequence between $g(I_{\sf init})$ and $g(I_{\sf fin})$ of length $3\ell$. We therefore conclude that there is a reconfiguration sequence of length $\ell$ between $I_{\sf init}$ and $I_{\sf fin}$ if and only if there is one of length $3\ell$ between $g(I_{\sf init})$ and $g(I_{\sf fin})$.
\end{proof}

\begin{lemma}\label{thm:varIS4}\label{lemma:lastIS}
There exists a pl-reduction $f$ from {\sc Partitioned TJ-Independent Set Reconfiguration} to {\sc TJ-Independent Set Reconfiguration}, such that for any instance $\mathcal{A}$ of {\sc Partitioned TJ-Independent Set Reconfiguration},  $\mathcal{A}$ admits a reconfiguration sequence of length $\ell$ if and only if $f(\mathcal{A})$ admits a reconfiguration sequence of length $3\ell$.
\end{lemma}
 \begin{proof}
Let $\mathcal{A}=(G,I_{\sf init},I_{\sf fin},k)$ be an instance of {\sc Partitioned TJ-Independent Set Reconfiguration} with $G=(V,E)$ and $P_1,\dots,P_k$ the token sets. 

The construction is similar to the one in the proof of Lemma \ref{thm:varIS3}.
We first create token gadgets.
Let $i\in [1,k]$. We create two copies of each vertex $v \in P_i$, called $v^a$ and $v^b$, and add the edges $v^aw^a$, $v^bw^b$ and $v^aw^b$ for all vertices $v\neq w$ in $P_i$. We also add the edges $v^aw^a$ if $vw\in E$. Moreover, we add a vertex $\delta_i$, which is connected to $v^b$ for all $v\in P_i$.  This forms the $i$th token gadget. Finally, we connect $\delta_i$ and $\delta_j$ for all $i,j\in[1,k]$. Let $G'$ be the resulting graph. 

We claim that any independent set $I'$ of $G'$ can contain at most two vertices of any token gadget. Let $i\in [1,k]$. If $I'$ contains $\delta_i$, then it cannot contain anything from $\{v^b:v\in P_i\}$. Since $I'$ can contain at most one vertex from $\{v^a:v\in P_i\}$ (which forms a clique), it intersects $I'$ in at most two vertices. In fact $I'$ can only contain two vertices from the $i$th token gadget if it contains $v^a$ and $v^b$, or $v^a$ and $\delta_i$, for some $v\in P_i$. Since we will have $2k$ tokens, exactly two tokens must be in each of the token gadgets. This models a choice of $v\in P_i$ for each $i\in [1,k]$. 
By enforcing that at most one $\delta_i$ can contain a token, we enforce that the vertex $v^a$ can be changed for $v\in P_i$ for only a single $i\in [1,k]$ at a time. 

The proof continues as in the proof of Lemma \ref{thm:varIS3}: each jump of the $i$th token from $v$ to $w$ in $G$ is modelled by three jumps in $G'$, namely $v^b \to \delta_i$, $v^a \to w^a$ and $\delta_i \to w^b$, and at least three jumps need to take place in order to move a token from $v^a$ to $w^a$ in $G'$.
\end{proof}


\subsection{Equivalences for Dominating Set}

\begin{lemma}\label{thm:varDS1}\label{lemma:firstDS}
There exists a pl-reduction $f$ from {\sc TJ-Dominating Set Reconfiguration} to {\sc TS-Dominating Set Reconfiguration}, such that for any instance $\mathcal{A}$ of {\sc TJ-Dominating Set Reconfiguration} with $k$ tokens,  $\mathcal{A}$ admits a reconfiguration sequence of length $\ell$ if and only if $f(\mathcal{A})$ admits a reconfiguration sequence of length $\ell+k+1$.
\end{lemma}
\begin{proof}
Let $\mathcal{A}=(G,D_{\sf init},D_{\sf fin},k)$ be an instance of  {\sc TJ-Dominating Set Reconfiguration} with $G=(V,E)$ and $k=|D_{\sf init}|$ tokens. We create an instance $f(\mathcal{A})$ of {\sc TS-Dominating Set Reconfiguration} with $k+1$ tokens. For each token $i\in [1,k]$ of $\mathcal{A}$, we first create the following gadget that models on which vertex of $G$ the token is. Let $V^i = \{v^i : v\in V\}$ be a copy of $V$ and let it induce an $n$-vertex clique. To certify that at least one token is in each $V^i$, we add vertices ${\sf gar}^i$ and ${\sf gar'}^i$, both connected to all vertices in $V^i$. If a token is placed on $v^i$, this corresponds to placing the $i$th token on vertex $v$. 

We want to prohibit the tokens to be on copies of the same vertex $v\in V$. To do this, we create a vertex $x^{i,j}_v$  for all $i,j\in [1,k]$ ($i\neq j$) and all $v \in V$, which we connect to $w^i$ and $w^j$ for all $w \in V\setminus\{v\}$. Assuming that exactly one token is in each $V^i$, this implies that not both $v^i$ and $v^j$ can be in the dominating set, because then $x^{i,j}_v$ is not dominated.

The chosen tokens should be a dominating set. To accomplish this, we add a vertex $v'$ for all $v\in V$, with edges $v'w^i$ for all $i\in[k]$ and all $w^i\in V^i$ such that $vw\in E$ or $v=w$. This results in a graph $G'$. 

There is still the issue for any dominating set in $G$, $k!$ different dominating sets in $G'$ correspond to it, which is an issue when defining the initial and final dominating sets. To solve this, we add $k+4$ more vertices $f^1,\dots,f^k$, $z_{\sf init}, z_{\sf fin}, z_{\sf gar}$ and $z'_{\sf gar}$. We add edges $f^i{\sf gar}^i$, $f^i{\sf gar'}^i$ and $f^iv^i$ for all $v\in D_{\sf fin}$ for all $i\in[1,k]$. This is such that a slide from $v^i$ to $f^i$ can only happen if $v \in D_{\sf fin}$. Furthermore we connect $z_{\sf gar}$ and $z'_{\sf gar}$ only to $z_{\sf init}$ and $z_{\sf fin}$ to ensure one token to always be on $z_{\sf init}$ or $z_{\sf fin}$. We add edges $f^iz_{\sf init}$ for all $i\in[1,k]$.  Finally, we connect $z_{\sf fin}$ to $z_{\sf init}$ and to all $v',v^1,\dots,v^k$ for all $v\in V$.  Call the resulting graph $G''$. 

Note that because of the `guardian' vertices, the pigeonhole principle implies that exactly one token is in the set $V^i\cup\{f^i\}$ for all $i\in[1,k]$ and exactly one token is on either $z_{\sf init}$ or $z_{\sf fin}$.

Let $D_{\sf init} = \{v_1,\dots,v_k\}$ be the initial dominating set of $\mathcal{A}$. 
We set $D'_{\sf init} = \{(v_i)^i : i\in[1,k]\} \cup \{z_{\sf init}\}$ and $D'_{\sf fin} = \{f^1,\dots,f^k,z_{\sf fin}\}$.

We omit the details why this construction works, and only give an informal explanation. A TJ-reconfigurations sequence $D_{\sf init} = D_0,D_1,\dots,D_{\ell}=D_{\sf fin}$ in $\mathcal{A}$ can be converted to a TS reconfiguration sequence $D'_{\sf init}=D'_0,D'_1,\dots,D'_{\ell}$ by numbering the tokens in $D_{\sf init}$ the way we did for $D'_{\sf init}$ and `tracking' their movements to the sequence, mimicking the movements within $G''$. Once we reach some set $D'_{\ell}$ that `corresponds' to $D_{\sf fin}$, we can move the tokens from $D'_\ell$ to $D'_{\sf fin}$ by $k+1$ further moves: first move $z_{\sf init}$ to $z_{\sf fin}$ and then the other tokens to $f^1,\dots,f^k$. Note that the first move can only happen if all $f^i$ are dominated, meaning that $D'_{\ell}$ must indeed correspond to $D_{\sf fin}$. 

This implies that there is a sequence in $\mathcal{A}$ of length $\ell$ if and only if there is one in $f(\mathcal{A})$ of length $l+k+1$.
\end{proof}

\begin{lemma} \label{thm:varDS2}
There exists a pl-reduction $f$ from {\sc TS-Dominating Set Reconfiguration} to {\sc Partitioned TS-Dominating Set Reconfiguration}, such that for any instance $\mathcal{A}$ of {\sc TS-Dominating Set Reconfiguration} with $k$ tokens, $\mathcal{A}$ admits a reconfiguration sequence of length $\ell$ if and only if $f(\mathcal{A})$ admits a reconfiguration sequence of length $\ell+k+1$.
\end{lemma}
\begin{proof}
Let $\mathcal{A}=(G,D_{\sf init},D_{\sf fin},k)$ be an instance of {\sc TS-Dominating Set Reconfiguration} with $G=(V,E)$. We create the following instance of {\sc Partitioned TS-Dominating Set Reconfiguration} with $k'=k+2$ tokens. Create $k$ copies of $G$, denoted by $G^1,\dots,G^k$. Each of these copies is a token set, meaning that each dominating set must contain exactly one of the vertices of $G^i$. The chosen vertex in $G^i$ models the choice of the $i$th token in $\mathcal{A}$. 

We want all tokens to be copies on different vertices. To accomplish this, we create a vertex $x_v^{i,j}$ for all $i,j\in[1,k]$, ($i\neq j$), and all $v\in V$ with edges $x_v^{i,j}w^i$ and $x_v^{i,j}w^j$ for all $w\in V\setminus\{v\}$. This ensures that no two tokens are on the same copy of a vertex: if both $v^i$ and $v^j$ are in the dominating set, then $x_v^{i,j}$ is not dominated. 

The chosen vertices should be a dominating set. To check this, we add a vertex $v'$ for all $v\in V$, with edges $v'w^i$ for all $i\in[1,k]$ and all $w\in V$ such that $vw\in E$ or $v=w$. Finally, we add three vertices $y_{\sf dom},y_{\sf gar},y_{\sf gar}'$ and add edges $y_{\sf dom}y_{\sf gar}$, $y_{\sf dom}y_{\sf gar}'$ and $y_{\sf dom}v_i$ for all $i\in[1,k]$, $v\in V$. We then create a token set $\{v'\}_{v \in V}\cup\{y_{\sf dom},y_{\sf gar},y_{\sf gar}'\}$. Now $y_{\sf dom}$ is in any dominating set because of $y_{\sf gar}$ and $y_{\sf gar}'$. Also, $y_{\sf dom}$ dominates all vertices in $G^1,\dots,G^k$. We choose to add $\{v'\}_{v\in V}$ to this token set, because the token sets should form a partition of the vertices in the instance $f(\mathcal{A})$. We call this graph $G'$.

We note that the $i$th token in $G$ can slide from $v$ to $w$ in $\mathcal{A}$ if and only if token $i$ in $f(\mathcal{A})$ can slide from $v_i$ to $w_i$. The proof now continues as in the proof of Lemma~\ref{thm:varDS1}: there is again a $k!$-to-one correspondence between dominating sets $D'$ of $G'$ and dominating sets $D$ of $G$ (for $D'$ of size $k+1$ in $G'$, we consider the set of $v\in V$ for which $v^i\in D'$ for some $i\in[1,k]$). We obtain $G''$ from $G'$ by adding new vertices $f^1,\dots,f^k,z_{\sf init},z_{\sf fin},z_{\sf gar},z'_{\sf gar}$ and the same edges as in the proof of Lemma~\ref{thm:varDS1}, creating a new token set for the vertices $z_{\sf init},z_{\sf fin},z_{\sf gar},z'_{\sf gar}$ and adding $f^i$ to the $i$th token set. The remainder of the analysis is analogous. 
\end{proof}

\begin{lemma} \label{thm:varDS3}
There exists a pl-reduction $f$ from {\sc Partitioned TS-Dominating Set Reconfiguration} to {\sc Partitioned TJ-Dominating Set Reconfiguration}, such that for any instance $\mathcal{A}$ of {\sc Partitioned TS-Dominating Set Reconfiguration}, $\mathcal{A}$ admits a reconfiguration sequence of length $\ell$ if and only if $f(\mathcal{A})$ admits a reconfiguration sequence of length $3\ell +1$.

\end{lemma}
\begin{proof}
Let $\mathcal{A}=(G,D_{\sf init},D_{\sf fin},k)$ be an instance of {\sc Partitioned TS-Dominating Set Reconfiguration} with $G=(V,E)$ and $P_1,\dots,P_k$ the token sets. We create $k$ token gadgets in $f(\mathcal{A})$, each modelling the choice of one token, as follows. Let $i\in [1,k]$. We create vertices $v^a$, $v^b$, $v^c$ and $v^d$ for all $v \in P_i$. We create token sets $\{v^a\}_{v \in P_i}$ and $\{v^c\}_{v \in P_i}$, meaning exactly one vertex of those sets should be in any dominating set.  
The following edges are created:
\begin{align*}
    v^av^b &\qquad\forall v\in V, &
    v^cv^d &\qquad\forall v\in V,\\
    v^aw^d &\qquad\forall v,w\in V, \, v\neq w,&
    v^cw^b &\qquad\forall v,w\in V,\, v\neq w.\\
    v^aw^a &\qquad\forall v,w\in V, \, v\neq w,&
    v^cw^c &\qquad\forall v,w\in V,\, v\neq w.
\end{align*}
The edges ensure that the two tokens of the token sets $\{v^a\}_{v \in P_i}$ and $\{v^c\}_{v \in P_i}$ can only together dominate all $w^b$ and $w^d$ vertices if they are on the copies of the same vertex. This implies that it is not possible to jump within these token sets, unless these is some other interaction with the gadget. For $i\in [1,k]$, we model the $i$th token being on vertex $v\in P_i$ (in the original graph) by a token on $v^a$. 


To model the sliding of the tokens in $\mathcal{A}$, we create vertices $\delta_{vw}$ for all $vw\in E(G[P_i])$ for all $i\in[1,k]$, and connect $\delta_{vw}$ to $v^b$, $w^b$, $v^d$ and $w^d$. We also create a `dummy' vertex $\delta_{\emptyset}$.
We place all $\delta_{\star}$ vertices within a single token set,  for a token that we call the \emph{edge token}. We connect all the vertices within this token set.

To verify that our choice of tokens induces a dominating set, we add another copy $v'$ for all $v\in V$, and connect $v'$ with $w^a$ for all $w\in V$ such that $vw\in E$ or $v=w$. Because all vertices should be part of a token set, we create one additional vertex, $z_{\sf dom}$, connected to no other vertex. We then have the token set $\{v'\}_{v \in V}\cup\{v^b\}_{v\in V}\cup\{v^d\}_{v\in V}\cup\{z_{\sf dom}\}$, where the token should always be on $z_{\sf dom}$. Let $G'$ be the resulting graph.

Let the initial and final dominating sets for $G'$ be \begin{align*}
D'_{\sf init} &= \{v^a,v^c\}_{v\in D_{\sf init}}\cup \{z_{\sf dom}\} \cup \{\delta_{\emptyset}\} \text{ and }\\
D'_{\sf fin} &= \{v^a,v^c\}_{v\in D_{\sf fin}}\cup \{z_{\sf dom}\} \cup \{\delta_{\emptyset}\}.
\end{align*}
We create an instance $f(\mathcal{A}) = (G',D'_{\sf init},D'_{\sf fin},2k+2)$ with the token sets as described above.
For any reconfiguration sequence $D_{\sf init} =D_0,D_1,\dots,D_\ell = D_{\sf fin}$ in $\mathcal{A}$ of length $\ell$, there is a reconfiguration sequence of length $3\ell+1$ in $f(\mathcal{A})$ of the form 
\[D_{\sf init}=g_1(D_0),g_2(D_0),g_3(D_0),\dots,g_1(D_{\ell})=D'_{\sf fin},\] 
where
\begin{align*}
    &g_1(D) =  \{v^a,v^c\}_{v\in D}\cup \{z_{\sf dom}\} \cup \{\delta_{x'y'}\}, &&\text{for } x'y' \text{ last slide-edge},\\
    &g_2(D) =  \{v^a,v^c\}_{v\in D}\cup \{z_{\sf dom}\} \cup \{\delta_{xy}\}, &&\text{for } xy \text{ next slide-edge},\\
    &g_3(D) =  (\{v^a,v^c\}_{v\in D}\setminus\{x^a\}\cup\{y^a\})\cup \{z_{\sf dom}\} \cup \{\delta_{xy}\}, &&\text{for } xy \text{ next slide-edge}.
\end{align*}
(We let the last slide-edge of $D_{\sf init}$ be the edge corresponding to the edge token in $D'_{\sf init}$, and the next slide-edge of $D_{\sf fin}$ be the edge corresponding to the edge token in $D'_{\sf fin}$.)

We say the $i$th token is `sliding' in $f(\mathcal{A})$ if there are tokens on $v^a$ and $w^c$ for $v\neq w$ elements of $P_i$.
Note that at most one token can be `sliding' at the same time in $f(\mathcal{A})$: the edge token can only move away from $\delta_{vw}$ if there are two tokens on $v^a$ and $v^c$, or two tokens on $w^a$ and $w^c$. Therefore, any minimal length reconfiguration sequence in $f(\mathcal{A})$ is of some length $3\ell +1$ (namely three steps per slide and one additional jump to get to $D'_{\sf fin}$). Any such reconfiguration sequence of $f(\mathcal{A})$ from $g_1(D_{\sf init})$ to $g_1(D_{\sf fin})$ must have a subsequence of length $\ell$ of distinct dominating sets $g_1(D_0),\dots,g_1({D_\ell})$ (where we take every third dominating set in the sequence). This gives a sequence of length $\ell$ in $\mathcal{A}$.
\end{proof}

\begin{lemma}\label{thm:varDS4}\label{lemma:lastDS}
There exists a pl-reduction $f$ from {\sc Partitioned TJ-Dominating Set Reconfiguration} to {\sc TJ-Dominating Set Reconfiguration}, such that for any instance $\mathcal{A}$ of {\sc Partitioned TJ-Dominating Set Reconfiguration}, $\mathcal{A}$ admits a reconfiguration sequence of length $\ell$ if and only if $f(\mathcal{A})$ admits a reconfiguration sequence of length $\ell$.
\end{lemma}
 \begin{proof}
 We keep the same number of tokens $k$.
 Let $P_1,\dots,P_k$ be the token sets of the {\sc Partitioned TJ-Dominating Set Reconfiguration} instance. For $i\in[1,k]$, we add vertices ${\sf gar}_i$ and ${\sf gar}_i'$, both connected to all vertices in $P_i$. This ensures that there is exactly one token in each of the sets $P_i$, modeling the token sets of $\mathcal{A}$.  As a consequence, only jumps within $P_i$ are allowed.  Therefore, any jump in $\mathcal{A}$ can be modelled by a jump in $f(\mathcal{A})$ and vice versa. 
 \end{proof}
 
 \subsection{Proof of Theorem \ref{thm:equivalences}}
All variants of \textsc{Independent Set Reconfiguration} and \textsc{Dominating Set Reconfiguration} are equivalent under pl-reductions by Lemmas~\ref{lemma:firstIS}--\ref{lemma:lastDS}. 

To see that these results also hold for the \emph{timed} reconfiguration variants, we take a closer look at the aforementioned reductions. 
For each reduction $f$ in one of the Lemmas~\ref{lemma:firstIS}--\ref{lemma:lastDS}, we gave a function $h(k,\ell)$ (linear in its two parameters) such that instance $\mathcal{A}$ with $k$ tokens has a reconfiguration sequence of length $\ell$ if and only if $f(\mathcal{A})$ has a reconfiguration sequence of length $h(k,\ell)$. 
Therefore the bound on the sequence length $\ell$ given in instances $\mathcal{A}$ of \textsc{Timed Independent Set Reconfiguration}, \textsc{Binary Independent Set Reconfiguration} or \textsc{Timed Independent Set Reconfiguration} (when moreover parameterized by the length of the sequence), only grow polynomially in $\ell$ and $k$ when reducing to an instance $f(\mathcal{A})$ of the same problem (but a different sliding/jumping variant).
The same arguments hold for the dominating set variants. This concludes the proof of Theorem~\ref{thm:equivalences}.

We note pl-reductions are a special case of fpt-reductions, as a space limit of $\mathcal{O}(g(k)+\log n)$ implies a runtime bound of $\mathcal{O}(2^{g(k)}\cdot n^c)$ for some constant $c$.

\section{XL-completeness of Reconfiguration without a Bound on the Sequence Length}
\label{sec:XLIS}

In this section, we give the complexity results for the case that the length of the sequence is not specified. For all
considered move rules, and both for the reconfiguration of independent sets and of dominating sets, we have XL-completeness.
A main ingredient is the fact that XL$=$XSL; we show membership by simulation on a Symmetric Turing Machine, and
hardness by a reduction from {\sc Accepting Log-Space Symmetric  Turing Machine}.

\subsection{Independent Set}
\label{section:is-xl}

The main step of our proof is to show that the following problem is XL-complete.

\begin{verse}
    {\sc Partitioned TS-Independent Set Reconfiguration}\\
    {\bf Given:} Graph $G=(V,E)$; independent sets $I_{\sf init},I_{\sf fin}$ of size $k$; a partition $V=\sqcup_{i=1}^k P_i$ of the vertex set.\\
    {\bf Parameter:} $k$.\\
    {\bf Question:} Does there exist a sequence $I_{\sf init}=I_0,I_1,\dots,I_T=I_{\sf fin}$ of independent sets of size $k$ for some $T$, with $|I_t\cap P_i|=1$ for all $t\in [0,T]$ and $i\in [1,k]$, such that for all $t\in [1,T]$, $I_{t}=I_{t-1}\setminus \{u\}\cup \{v\}$ for some $uv\in E(G)$ with $u\in I_{t-1}$ and $v \not \in  I_{t-1}$? 
\end{verse}
Theorem~\ref{thm:equivalences} then implies the XL-completeness results for the other variants of {\sc Independent Set Reconfiguration}.


\begin{theorem}
{\sc Partitioned TS-Independent Set Reconfiguration} is $\XL$-complete. \label{thm:xl_IS} 
\end{theorem}
\begin{proof}
By Theorem \ref{theorem:xsl=xl}, it suffices to give pl-reductions to and from {\sc Accepting Log-Space Symmetric Turing Machine}.

The problem is in XL (=XSL) as it can be simulated with a Symmetric Turing Machine with $\mathcal{O}(k \log n)$ space as follows. We store the current independent set of size $k$ on the work tape, which takes about $k\cdot \log n$ bits space. We use the transitions of the STM to model the changes of one of the vertices in the independent set. 
For all vertices $u,v\in V$, we have a sequence of states and transitions that allows you to remove $u$ and add $v$ to the independent set currently stored on the work tape, if the following assumptions are met:  $u\in I$, $v\not \in I$, $uv\in E(G)$, $u,v$ are part of the same token set (part of the partition) and $I'$ is an independent set. This gives a total of $O(n^2k^2)$ states. There is one accepting state, reachable via a sequence of states and transitions that verifies that the current independent set is the final independent set.
All transitions are symmetric.

We prove the problem to be XL-hard by giving a reduction from {\sc Accepting Log-Space Symmetric  Turing Machine}. Let $\mathcal{M}=(\mathcal{S},\Sigma,\mathcal{T},s_{\sf start},\mathcal{A})$ be the STM of a given instance, with $\mathcal{A} = \{s_{\sf acc}\}$, $\Sigma = [1,n]$ and a work tape of $k$ cells. We may assume that $\mathcal{M}$ only accepts if the symbol $1$ is on every cell of the work tape and the head is at the first position. 
This can be done by creating a new accepting state and adding $\mathcal{O}(k)$ transitions from $s_{\sf acc}$ to this new state, which set only $1$'s on the work tape and move the head to the first position. We create an instance $\Gamma$ of {\sc Partitioned TS-Independent Set Reconfiguration} with $k' = k+1$ tokens. These tokens will simulate the configuration of $\mathcal{M}$: $k$ tape-tokens modeling the work tape cells and one state token describing the current state and tape head position. 

\subparagraph*{Tape gadgets.} For each work tape cell $i \in [1,k]$, we create a \emph{tape gadget} consisting of $n+1$ vertices as follows. We add a vertex $v^i_\sigma$ for all $\sigma \in \Sigma = [1,n]$ and a vertex $y^i$, connected to $v^i_\sigma$ for all $\sigma \in \Sigma$. The vertices in a tape gadget form a token set (a part of the partition, i.e. exactly one of these $n+1$ vertices is in the independent set at any given time). The symbol $\sigma$ that is on the $i$th work tape cell of $\mathcal{M}$ is simulated by which $v^i_\sigma$ is in the independent set. 

\subparagraph*{State vertices.}
The last token set is the set of all transition and state vertices (defined below), meaning that exactly one of these vertices is in the independent set at any given time. The token of this set (called the `state token') simulates the state of $\mathcal{M}$, the position of the head, and the transition $\mathcal{M}$ takes. 

We create a vertex $p^{i}$ for each state $p\in \mathcal{S}$ and all head positions $i \in [1,k]$. We add edges $p^{i}y^{i'}$ for all $i' \in [1,k]$. These vertices will simulate the current state of $\mathcal{M}$ and the position $i$ of the tape head. 

\subparagraph*{Transition vertices.}
To go from one state vertex to another, we create a path of three transition vertices, to checking whether the work tape agrees with the transition before and afterwards, and one allowing moving some tokens of the tape gadget. 
To control when we can move tokens in the tape gadgets, we put edges between $y^{i}$ and all state and transition vertices (for each $i \in [1,k]$), unless specified otherwise. We further outline which edges are present below, and give an example in Figure~\ref{fig:XLproofIS}.

\begin{figure}[h]
\centering
\scalebox{.9}{
\begin{tikzpicture}[scale=.9,
dot/.style = {circle, draw, minimum size=#1,
              inner sep=0pt, outer sep=0pt},
dot/.default = 1 cm,
dot2/.style = {circle, draw, minimum size=#1,
              inner sep=0pt, outer sep=0pt},
dot2/.default = .7 cm]

    \node[dot] (Sp) at (1,0) {$p^{i}$};
    \node[dot] (Sq) at (9,0) {$q^{i+1}$};
    \node[dot] (Tab) at (3,0) {$\tau_{ab}^{i}$};
    \node[dot] (Tsh) at (5,0) {$\tau_{\sf shift}^{i}$};
    \node[dot] (Tcd) at (7,0) {$\tau_{cd}^{i}$};

    \draw[thick] (Sp) to (Tab);
    \draw[thick] (Tab) to (Tsh);
    \draw[thick] (Tsh) to (Tcd);
    \draw[thick] (Tcd) to (Sq);
    
    \node[dot2] (a1) at (0,7) {\small$v^i_a$};
    \node[dot2] (b1) at (0,6) {\small$v^i_b$};
    \node[dot2] (c1) at (0,5) {\small$v^i_c$};
    \node[dot2] (d1) at (0,4) {\small$v^i_d$};

    \node[dot2] (a2) at (10,7) {\small $v^{i+1}_a$};
    \node[dot2] (b2) at (10,6) {\small$v^{i+1}_b$};
    \node[dot2] (c2) at (10,5) {\small$v^{i+1}_c$};
    \node[dot2] (d2) at (10,4) {\small$v^{i+1}_d$};
    
    \node[dot2] (y1) at (-1,3) {\small$y^i$};
    \node[dot2] (y2) at (11,3) {\small$y^{i+1}$};
    
    \draw[thick] (a1) to [bend right=20] (y1);
    \draw[thick] (b1) to [bend right=20](y1);
    \draw[thick] (c1) to [bend right=20](y1);
    \draw[thick] (d1) to [bend right=20](y1);
    \draw[thick] (a2) to [bend left=20](y2);
    \draw[thick] (b2) to [bend left=20](y2);
    \draw[thick] (c2) to [bend left=20](y2);
    \draw[thick] (d2) to [bend left=20](y2);
    
    \draw[red] (y1) to [bend left=10](Sp);
    \draw[red] (y1) to [bend left=10](Tab);
    \draw[red] (y1) to [bend left=10](Tcd);
    \draw[red] (y1) to [bend left=10](Sq);
    
    \draw[red] (y2) to [bend right=10](Sp);
    \draw[red] (y2) to [bend right=10](Tab);
    \draw[red] (y2) to [bend right=10](Tcd);
    \draw[red] (y2) to [bend right=10](Sq);
    
    \draw[blue] (b1) to [bend left=20](Tab);
    \draw[blue] (c1) to [bend left=20](Tab);
    \draw[blue] (d1) to [bend left=20](Tab);
    \draw[blue] (a2) to [bend right=20](Tab);
    \draw[blue] (c2) to [bend right=20](Tab);
    \draw[blue] (d2) to [bend right=20](Tab);
    
    \draw[green!60!black] (a1) to [bend left=20](Tcd);
    \draw[green!60!black] (b1) to [bend left=20](Tcd);
    \draw[green!60!black] (d1) to [bend left=20](Tcd);
    \draw[green!60!black] (a2) to [bend right=20](Tcd);
    \draw[green!60!black] (b2) to [bend right=20](Tcd);
    \draw[green!60!black] (c2) to [bend right=20](Tcd);
\end{tikzpicture}}
\caption{Sketch of part of the construction of Theorem~\ref{thm:xl_IS}. Given are the two tape gadgets for positions $i$ and $i+1$, two state vertices and a transition path for transition $\tau=(p,(ab,1,cd),q)$.}
\label{fig:XLproofIS}
\end{figure}

Recall that we may assume that head always moves left or right. Consider first a transition $\tau\in \mathcal{T}$ that moves the head to the right, say $\tau=(p,(ab,1,cd),q)$. 
For all $i\in [1,k-1]$, we create a path between state vertices $p^{i}$ and $q^{i+1}$ consisting of three `transition' vertices: $\tau^{i}_{ab}$, $\tau^{i}_{\mathsf{shift}}$ and $\tau^{i}_{cd}$.
In order to ensure a token can only be on $\tau^{i}_{ab}$ when the token on the $i$th tape gadget represents the symbol $a$, we add edges from $\tau^{i}_{ab}$ to $v^{i}_\sigma$ for all $\sigma \in \Sigma\setminus \{a\}$. Similarly, edges to $v^{i+1}_\sigma$ are added for all $\sigma\in \Sigma \setminus \{b\}$, as well as edges between $\tau^{i}_{cd}$ and all $v^{i}_\sigma$ and $v^{i+1}_\sigma$ except for $v^i_c$ and $v^{i+1}_d$.
When the token is on the shift vertex $\tau^{i}_{\mathsf{shift}}$, the independent set is  allowed to change the token in the $i$th and $(i+1)$th tape gadget. Therefore, we remove the edges between $y^i$ and $y^{i+1}$ and $\tau^{i}_{\mathsf{shift}}$.  

Note that the constructed paths also handle transitions which move the tape head to the left, as we can transverse the constructed paths in both directions.  We omit the details.


\subparagraph*{Initial and final independent sets.}
Recall that $s_{\sf start}$ is the starting state of $\mathcal{M}$. We let the initial independent set be $I_{\sf init}= \{s_{\sf start}^{1}\}\cup \left(\bigcup_{i\in [1,k]} \{v^i_{1}\}\right)$, corresponding to the initial configuration of $\mathcal{M}$.
Let the final independent set be $I_{\sf{fin}}= \{s_{\sf acc}^{1}\} \cup \left(\bigcup_{i \in [1,k]} \{v^i_1\}\right) $, where $s_{\sf acc}$ is the accepting state of $\mathcal{M}$.  We note that both $I_{\sf init}$ and $I_{\sf fin}$ are independent sets. \\

Let $\Gamma$ be the created instance of {\sc Partitioned TS-Independent Set Reconfiguration}. 
We prove that $\Gamma$ is a yes-instance is and only if $\mathcal{M}$ accepts. 
We use the following function, hinting at the equivalence between configurations of $\mathcal{M}$ and certain independent sets of $\Gamma$.
\begin{definition}
Let $C = (p,i,\sigma_1,\dots,\sigma_k)$ be a configuration of $\mathcal{M}$. Then let $I(C)$ be the corresponding independent set in $\Gamma$:
$$I(C) = \{p^{i}\} \cup \left( \bigcup_{j=1}^k  \left \{v^{j}_{\sigma_{j}} \right\} \right) $$
\end{definition}
 
\begin{claim}
$\Gamma$ is a yes-instance if $\mathcal{M}$ accepts.
\end{claim}
\begin{proof}
Let $C_1,\dots,C_\ell$ be the sequence of configurations such that $\mathcal{M}$ accepts. We note that $I(C_1) = I_{\sf init}$. For each $t \in [1,\ell-1]$, we do the following. 
Let $C_t = (p,i,\sigma_1,\dots,\sigma_k)$. Let $\tau$ be the transition $\mathcal{M}$ takes to the next configuration. Assume $\tau=(p,(ab,1,cd),q)$, the case where $\tau$ moves the head to the left will be discussed later, but is similar.
Take the following sequence of independent sets, where $I(C_{t}) = I_0$ is the current independent set:
\begin{align*}
    I_1 &= I_0\setminus \left \{p^{i} \right\} \cup \left \{\tau^{i}_{ab}\right\} &    I_5 &= I_4\setminus \left \{v^{i+1}_b\right\} \cup \left \{y^{i+1}\right\} \\
    I_2 &= I_1\setminus \left \{\tau^{i}_{ab}\right\} \cup \left \{\tau^{i}_{\mathsf{shift}}\right\} &    I_6 &= I_5\setminus \left \{y^{i+1}\right\} \cup \left \{v^{i+1}_d \right\}  \\
    I_3 &= I_2\setminus \left \{v^i_a \right\}\cup \left \{y^i \right\}&I_7 &= I_6\setminus \left \{\tau^{i}_{\sf{shift}} \right\}\cup \left \{\tau^{i}_{cd}\right\}\\
    I_4 &= I_3\setminus \left \{y^i \right\}\cup \left \{v^i_c\right\} & I_8 &= I_7\setminus \left \{\tau^{i}_{cd} \right\} \cup \left \{q^{i+1}\right\}
\end{align*} 
Notice that this sequence of independent sets is allowed, as all sets are independent, each token stays within its token set and each next independent set is a slide away from its previous. Also, we see that $I(C_{t+1}) = I_8$. 
For transition $\tau=(p,(ab,-1,cd),q)$, where the tape head moves to the left, we do the following. Let $\tau^{-1} = (q,(cd,1,ab),p)$ be the inverse. We take the sequence that belongs to $\tau^{-1}$ backwards, i.e. if $I_0,\dots,I_8$ was the sequence of independent sets as described for $\tau^{-1}$, then take the sequence $I_8,\dots,I_0$.

We note that $I(C_\ell) = I_{\sf fin}$ is the final independent set, as we assumed the machine only to accept with $\sigma_i =1$ for all $i\in[1,k]$ and the head at the first position. Therefore, we find that this created sequence of independent sets is a solution to $\Gamma$. 
\end{proof}

We now prove the other direction.
\begin{claim}
$\mathcal{M}$ accepts if $\Gamma$ is a yes-instance.
\end{claim}
\begin{proof}
Let $I_{\sf init} = I_1,\dots,I_{\ell-1},I_\ell = I_{\sf fin}$ be the sequence of independent sets that is a solution to $\Gamma$. We assume this sequence to be minimal, implying that no independent set can occur twice. 

The state token should always be on either a state or transition vertex, because of its token set. Let $I'_1,\dots,I'_{\ell'}$ be the subsequence of $I_1,\dots,I_\ell$ of independent sets that include a state vertex. We will prove that the configurations of $\mathcal{M}$, simulated by this subsequence, is a sequence of configurations that leads to the accepting state $s_{\sf acc}$. To do this, first we note some general facts about $I_t$ for $t \in [1,\ell]$.

If the state token of $I_t$ is on a state vertex $p^{i}$, then $I_{t+1}$ slides the state token to a neighbor of $p^{i}$. This is because all $y^{i'}$ for $i'\in[1,k]$ are neighbors of $p^{i}$, hence the tokens in the tape gadgets are on some $v^{i'}_\sigma$ and cannot move. The same holds for transition vertices of the form $\tau^{i}_{ab}$. If $\tau^{i}_{\sf shift} \in I_t$, then $y^i$ and $y^{i+1}$ are not neighbors of the state token. Therefore, the $i$th and $i+1$th tape gadgets token can now slide.
If $\tau^{i}_{ab} \in I_t$, then $v^i_{a}\in I_t$ and $v^{i+1}_{b} \in I_t$. This is because all other vertices of the $i$th and $i+1$th tape gadgets are neighbors of $\tau^{i}_{ab}$. 

Recall that $I'_1,\dots,I'_{\ell'}$ is the sequence of independent sets with the state token on a state vertex. For any $I'_t$ with $t\in [1,\ell']$, let $C_t$ be the unique configuration of $\mathcal{M}$ such that $I(C_t) = I'_t$. We prove that $C_1,\dots,C_{\ell'}$ is an allowed sequence of configurations for $\mathcal{M}$. Note that this implies that $\mathcal{M}$ accepts as $C_{\ell'}$ is the accepting configuration.

We fix $t \in [1,\ell']$ and focus on the transition between $C_t$ and $C_{t+1}$. Let $A_1,\dots,A_R$ be the sequence of independent sets in the solution of $\Gamma$, that are visited between $I'_t$ and $I'_{t+1}$. By definition of $I'_t$ and $I'_{t+1}$, $A_r$ does not contain a state vertex for all $r\in [1,R]$, therefore each $A_r$ must have its state token on a transition vertex. Each such transition vertex corresponds to the same transition $\tau=(p,\Delta,q)$, as this is the only path the state token can take. We assume that $\Delta= (ab,1,cd)$, the case $\Delta=(ab,-1,cd)$ can be proved with similar arguments. The set $A_1$ contains transition vertex $\tau^{i}_{ab}$ and therefore $I'_{t}$ contains $v^i_a$ and $v^{i+1}_b$. Also, $A_R$ contains $\tau^{i}_{cd}$, implying that $v^i_{c}, v^{i+1}_d \in I'_{t+1}$. We note that $A_2,\dots,A_{R-1}$ must contain $\tau^{i}_{\sf shift}$: only the $i$th and $i+1$th tape gadget tokens can shift when the state token is on $\tau^{i}_{\sf shift}$. So if the state token would be on $\tau^{i}_{ab}$ or $\tau^{i}_{cd}$ twice in $A_1,\dots,A_R$, the independent sets would be equal, contradicting the minimal length of the sequence. 

Combining this all, we conclude that if $I(C_t)=I'_t$ and $I(C_{t+1})=I'_{t+1}$, there is an allowed sequence of independent set, traversing the path belonging to a transition $\tau=(p,(ab,1,cd),q)$. Therefore, $I_{t+1} = I_{t}\setminus \{v^i_a,v^{i+1}_b,p^{i}\}\cup\{v^i_c,v^{i+1}_d,q^{i+1}\}$ and we are allowed to take transition $\tau$ from $C_t$ to end up in configuration $C_{t+1}$. 
\end{proof}

Hence, $\Gamma$ is a yes instance if and only if $\mathcal{M}$ accepts and we find that the given reduction is correct. This concludes the proof of Theorem~\ref{thm:xl_IS}. 
\end{proof}

With Theorem~\ref{thm:equivalences}, we can conclude the following.

\begin{corollary}
{\sc TS-Independent Set Reconfiguration} and {\sc TJ-Independent Set Reconfiguration} are XL-complete.
\end{corollary}

\subsection{XL-completeness of Dominating Set Reconfiguration}
\label{section:ds-xl}
We now prove {\sc Partitioned TS-Dominating Set Reconfiguration} to be XL-complete. Note that Theorem~\ref{thm:equivalences} implies the same complexity for all mentioned variants Dominating Set Reconfiguration of (partitioned) token jumping/sliding.
\begin{theorem}
{\sc Partitioned TS-Dominating Set Reconfiguration} is \emph{XL}-complete.
\end{theorem}
\begin{proof}
The problem is in XL by similar arguments as with {\sc Partitioned TS-Independent Set Reconfiguration}  as it can be simulated with an STM using $\mathcal{O}(k \log n)$ space.

We prove the problem to be XL-hard by giving a reduction from {\sc Accepting Log-Space Symmetric Turing Machine}. The reduction is almost identical to the reduction in Theorem~\ref{thm:xl_IS}, therefore we will not discuss correctness of the construction and only give a short description.

Let $\mathcal{M}=(\mathcal{S},\Sigma,\mathcal{T},s_{\sf start},\mathcal{A})$ be the STM of a given instance, with $\mathcal{A} = \{s_{\sf acc}\}$, $\Sigma = [1,n]$ and work tape of $k$ cells. We create an instance $\Gamma$ of {\sc Partitioned TS-Dominating Set Reconfiguration} with $k' = k+2$ tokens, so one token more than in the reduction of Theorem~\ref{fig:XLproofIS}. 

\subparagraph*{Tape gadgets.} For each work tape cell $i\in[1,k]$, we create a \emph{tape gadget} consisting of $n+2$ vertices: $\{v^i_1,\dots,v^i_n,x^i,y^i\}$. These vertices form a token set. We add edges $x^iv^i_\sigma$ and $y^iv^i_\sigma$ for all $\sigma \in \Sigma=[1,n]$. 

\subparagraph*{State vertices.}
We create a state vertex $p^{i}$ for each state $p\in \mathcal{S}$ and all head positions $i \in [1,k]$. We add edges $p^{i}v^{i'}_\sigma$ for all $i' \in [1,k]$ and all $\sigma \in \Sigma$. 

\subparagraph*{Transition vertices.}
As in the proof of Theorem~\ref{fig:XLproofIS}, we create a path of three transition vertices for each allowed transition between two state vertices. We add edges between all $v^i_\sigma$ vertices and all state and transition vertices (for all $i\in[1,k]$), unless specified otherwise.

In order to ensure a token can only be on $\tau^{i}_{ab}$ when the token of the $i$th tape gadget is on the vertex that represents the symbol $a$, we remove the edge $\tau^{i}_{ab}v^i_a$. For the $i$th tape gadget, now all vertices except for $x^i$, $y^i$ and $v^i_a$ are dominated by $\tau^{i}_{ab}$, hence the token must be on $v^i_a$. Similarly, we remove the edge $\tau^{i}_{ab}v^{i+1}_b$, as well as edges $\tau^{i}_{cd}v^i_c$ and $\tau^{i}_{cd}v^{i+1}_d$. When the token is on the shift vertex $\tau^{i}_{\sf shift}$, the dominating set should be allowed to change the tokens in the $i$th and $(i+1)$th tape gadgets. Therefore, we add edges $\tau^{i}_{\sf shift}x^i$ and $\tau^{i}_{\sf shift}x^{i+1}$, such that the token is able to slide to $y^i$ and $y^{i+1}$.

One of the tokens, called the `state token', gets the set of all transition and state vertices as its token set. The position of the state token simulates the state of $\mathcal{M}$, the position of the head, and the transitions $\mathcal{M}$ takes. 

\subparagraph*{Dominating vertex.}
We create a token set, consisting out of one vertex $z_{\sf dom}$. This vertex is connected to all state and transition vertices. As $z_{\sf dom}$ is, by construction, in any dominating set, it allows us to assume that all state and transition vertices are dominated.

\subparagraph*{Initial and final dominating sets}
Recall that $s_{\sf start}$ is the starting state of $\mathcal{M}$. We let the initial dominating set be $D_{\sf init}=\{ s_{\sf start}^1\} \cup \{z_{\sf dom}\}\cup \left(\bigcup_{i=1}^k \{v^i_{1}\}\right) $, corresponding to the initial configuration of $\mathcal{M}$. 
Let the final dominating set be $D_{\sf{fin}}= \{s_{\sf acc}^{1}\} \cup \{z_{\sf dom}\}\cup\left(\bigcup_{i=1}^k \{v^i_1\}\right) $, where $s_{\sf acc}$ is the accepting state of $\mathcal{M}$. We note that $D_{\sf init}$ and $D_{\sf fin}$ are dominating sets.
 
Let $\Gamma$ be the created instance of {\sc Partitioned TS-Dominating Set Reconfiguration}. Because of almost identical arguments as in the proof of Theorem~\ref{thm:xl_IS}, $\Gamma$ is a yes-instance if and only if $\mathcal{M}$ accepts.
\end{proof}

\begin{corollary}
{\sc TS-Dominating Set Reconfiguration} and {\sc TJ-Dominating Set Reconfiguration} are XL-complete.
\end{corollary}

\section{XNL-completeness of Binary Timed Reconfiguration}
\label{sec:XNL-c}
In this section, we show XNL-completeness of \textsc{Binary Timed Independent Set Reconfiguration} (Section~\ref{section:is-xnl}),
and \textsc{Binary Timed Dominating Set Reconfiguration} (Section~\ref{section:ds-xnl}).
Before giving the main results, we derive an intermediate result, that gives XNL-completeness of a special version of
\textsc{Satisfiability}. The result possibly has independent interest, and could be used for other XNL-hardness proofs.

\subsection{Long Chained Satisfiability}
\label{section:longchain}

In~\cite{BodlaenderGNS21}, `chained variants' of \textsc{Satisfiability} were introduced, and proved to be XNLP-complete.
Such problems can be useful to give XNLP-hardness proofs for problems from various settings. Here, we introduce
a \emph{long chain} variant of one of these problems. We remark that most variants of \textsc{Chained Satisfiability}
given in \cite{BodlaenderGNS21} have a `long chain' variant that is XNL-complete; and proving such XNL-completeness can be done
in a similar way. Here, we only present the form that is needed for the results proven later in this section.

\begin{verse}
\longchain\\
{\bf Input}: Integers $k,q,r\in \N$
with $r$ given in binary and $r\leq q^k$; Boolean formula $F$, which is an expression on $2q$ positive variables and in conjunctive normal form; a partition of $[1,q]$ into $k$ parts $P^1,\dots,P^k$. \\
{\bf Parameter}: $k$.\\
{\bf Question}: Do there exist variables $x_j^{(t)}$ for $t\in [1,r]$ and $j\in [1,q]$, such that we can satisfy the formula 
$$ 
\bigwedge_{1\leq t\leq r-1} F(x_1^{(t)},\dots,x^{(t)}_q,x_1^{(t+1)},\dots,x^{(t+1)}_q)
$$
by setting, for $i\in [1,k]$ and $t\in [1,r]$,  exactly one variable from the set $\{x_j^{(t)} : j\in P_i\}$ to true and all others to false?
\end{verse} 

\begin{theorem}
\label{thm:xnl_CNF}
\longchain ~is $\XNL$-complete.
\end{theorem}

\begin{proof}
We first show that the problem is in XNL: an instance of \longchain~can be simulated by a Nondeterministic Turing Machine with a work tape that can contain $3k$ symbols from an alphabet of size $q$. Of these, $k$ symbols are used to represent our `current time', an integer $t\in [1,r]$ (where $r\leq q^k$). For each $i \in [1,k]$, we use two symbols from $[1,q]$ to represent which variables from $\{x_j^{(t)}: j\in P_i\}$ and $\{x_j^{(t+1)}: j\in P_i\}$ are set to true, where $t$ is the symbol representing the `current time'. Finally, we create the transitions in such a way that if $F(x_1^{(t)},\dots,x^{(t)}_q,x_1^{(t+1)},\dots,x^{(t+1)}_q)$ is true, the `current time' goes from $t$ to $t+1$ (until we reach $r$, which brings us to the accepting state) and the $k$ corresponding symbols are renewed. If the formula is false, we transition to the rejecting state. 

Suppose now that we are given an instance of 
{\sc Accepting Log-Space Nondeterministic Turing Machine}.
Let $\mathcal{M}=(\mathcal{S},\Sigma,\mathcal{T},s_{\sf start},\mathcal{A})$ be the NTM of a given instance, with $\mathcal{A} = \{s_{\sf acc}\}$, $\Sigma = [1,n]$ and a work tape of $k$ cells. We may again assume that $\mathcal{M}$ only accepts if the symbol $1$ is in every cell of the work tape and the head is at the first cell.

Recall that \emph{configurations} of an NTM $\mathcal{M}$ are $k+2$ tuples describing the state, work tape and head position of $\mathcal{M}$ (see Definition~\ref{def:configuration}). 

There are at most $r=k|\mathcal{S}||\Sigma|^k$ different configurations of $\mathcal{M}$. Any shortest path from the stating configuration to an accepting configuration therefore has length at most $r$ and so we do not have to explore any simulations of the machines which take more transitions than that. 

We create four sets of variables, which `simulate the configuration of the machine':
\begin{itemize}
    \item $h_{t,p}$ for $t\in [1,r]$ and $p\in [1,k]$ represents the position $p$ of the tape head at time $t$;
    \item $z_{t,s}$ for $t\in [1,r]$ and $s\in \mathcal{S}$ represents the state $s$ of the machine at time $t$;
    \item $w_{t,p,\sigma}$ for $t\in [1,r]$, $p\in [1,k]$, $\sigma\in \Sigma$ represents that position $p$ of the work tape has symbol $\sigma$ at time $t$;
    \item $d_{t,\tau}$ for $t\in [1,r]$ and $\tau \in \mathcal{T}$ represents that the machine employs transition $\tau$ from time $t-1$ to $t$.
\end{itemize}
For each $t\in [1,r]$, the machine has exactly one state, the head one position, each work tape position one symbol and the machine takes exactly one transition. We ensure this by partitioning the variables appropriately. Namely, set $k'=k+3$ to be the number of parts and $q=k+|\mathcal{S}|+k|\Sigma|+|\mathcal{T}|$ the number of variables involved. For each $t\in [1,r]$, we create the parts
    \begin{align*}
        &\{ h_{t,p}: p\in [1,k]\}, \\
        &\{z_{t,s}:s\in \mathcal{S}\}, \\
        &\{ w_{t,p,\sigma}: \sigma\in \Sigma\} ~\text{ for all }p\in [1,k],\\
        &\{ d_{t,\tau}:\tau\in \mathcal{T}\}.
    \end{align*}
  This defines the partition of $[1,q]$. 
We create the following formulas which verify that the variables behave as we wish.
\begin{itemize}
    \item When a transition $\tau \in \mathcal{T}$ takes place from time $t-1$ to time $t$, then at time $t$ the work tape, state and head must take the values specified by the transition (this may depend on the current position of the head). For each $t\in [1,r]$ and transition $\tau\in \mathcal{T}$ that changes state $s$ to $s'$ and symbol $a$ to $b$ and moves the head to the right, we add the formulas
    \begin{align*}
&\neg d_{t,\tau} \vee z_{t,s'}, \\ 
    &\neg d_{t,\tau} \vee \neg h_{t-1,p} \vee w_{t,p,b} \text{ for each }p\in [1,k]\text{, and} \\  
&\neg d_{t,\tau} \vee \neg h_{t-1,p} \vee h_{t,p+1} \text{ for each }p\in [1,k]. 
\end{align*}
The first formula states that when $\tau$ is invoked from time $t-1$ to time $t$, the state becomes $s'$ at time $t$. The second and third ensure that if the head is on position $p$ at time $t-1$ when $\tau$ is invoked, then the symbol on position $p$ becomes $b$ at time $t$ and the head is in position $p+1$ at time $t$.
Transitions that move the head to the left or leave it on the same place are handled analogously.
\item At time $t-1$, the `initial conditions' of the transition that takes place need to be true.  For each $t\in [1,r]$ and transition $\tau\in \mathcal{T}$ that changes state $s$ to $s'$ and symbol $a$ to $b$, we add the formulas
    \begin{align*}
&\neg d_{t,\tau} \vee z_{t-1,s}, \text{ and } \\ 
    &\neg d_{t,\tau} \vee \neg h_{t-1,p} \vee w_{t-1,p,a} \text{ for each $p\in [1,k]$.}
\end{align*}
\item
Only the work tape position that the tape head is on can change. For each $t\in[1,r]$, and distinct head positions $p,p'\in [1,k]$, we add the formulas
\begin{align*}
    & \neg h_{t-1,p} \vee \neg w_{t-1,p',\sigma} \vee w_{t,p',\sigma} \text{ for all } \sigma\in\Sigma.
\end{align*}
This formula states if at time $t-1$, the head is at position $p$ and the work tape has symbol $\sigma$ at position $p'\neq p$, then the work tape still has symbol $\sigma$ at position $p'$ at time $t$. 
\item For the starting state $s_{\sf start}\in \mathcal{S}$ of the machine, we add a single formula of the form $z_{0,s_{\sf start}}$. We similarly initialise the work tape and head position appropriately. Finally, we add a transition from the accepting state $s_{\sf acc}$ to itself, and check whether at time $r$ we are in $s_{\sf acc}$ by adding the formula $z_{r,s_{\sf acc}}$. 
\end{itemize}
We need two further adjustments in order to have an instance of \longchain. 
First of all, we want to have a single formula $F$ that is used for all variable sets, but currently there are formulas $F_0$ and $F_2$ that we only wish to be true for the initial and final set of variables respectively (namely, those describing the initial and final configurations of the machine).
To obtain this, we `simulate a binary counter'.

Let us assume for convenience that $R=\log_k(r+1)$ is an integer.
A number $t\in [0,r]$ can be represented `in base $k$' as
\[
t=\sum_{j=0}^{R-1} b_{j} k^j
\]
with $b_j=b_j(t)\in [0,k-1]$. (When $R$ is not an integer, `the last' $b_j$ will have a smaller domain.)  We increase the number of parts $k'$ by $k$ and the number of variables $q$ by $Rk$. A new variable $b_{t,j,i}$ (with $j\in [0,R-1]$ and $i\in [0,k-1]$) models whether $b_j(t)=i$. 

For each $t\in [1,r]$, $\{b_{t,j,i}:i\in [0,k-1]\}$ forms a part. For all clauses $C_0$ in $F_0$ and $C_2$ in $F_2$, we add the two formulas
\begin{align*}
&\left(\bigvee_{j=0}^{R-1} \neg b_{t,j,0} \right )\vee C_0,\\
&\left(\bigvee_{j=0}^{R-1} \neg b_{t,j,k-1} \right )\vee C_2.
\end{align*}
We now enforce `the counter' to move up by one in each time step, such that if there is a satisfying assignment, $\wedge_{j=0}^R b_{t,j,0}$ holds if and only if $t=0$ and $\wedge_{j=0}^R \neg b_{t,j,k-1}$ holds if and only if $t=r$. This is done by adding the following formulas for all $j\in [1,R-1]$ and $i\in [0,k-1]$
\begin{align*}
    & \neg b_{t,0,i} \vee b_{t+1,0,i+1 \bmod k},\\
    &(\neg b_{t,j,i} \vee  b_{t,j-1,k-1}) \vee b_{t+1,j,i},\\
    &(\neg b_{t,j,i} \vee \neg b_{t,j-1,k-1}) \vee b_{t+1,j,i+1\bmod k}, \\
    & \left(\bigvee_{j=0}^{R-1} \neg b_{t,j,k-1}\right) \vee \left(\bigvee_{j=0}^{R-1 }\neg b_{t+1,j,0}\right).
\end{align*}
The last formula ensures that we can never let the counter go from $r$ to $0$, which ensures that the only way to move up the counter $r+1$ times is to start at $0$ for $t=0$ and end at $r$ at $t=r$.

\smallskip

Secondly, the occurrence of all literals must be positive. We can ensure this by replacing each occurrence of a negative literal by a requirement that one of the other variables in its part must be true, e.g. $\vee_{s'\neq s}z_{t,s}$ has the same logical meaning as $\neg z_{t,s}$ since exactly one variable from $\{z_{t,s'}:s'\in \mathcal{S}\}$ is known to be true.
\end{proof}

\subsection{Binary Timed Independent Set Reconfiguration}
\label{section:btisr}
\label{section:is-xnl}

We start by showing XNL-completeness for the following problem.

\begin{verse}
    {\sc Binary Timed Partitioned TS-Independent Set Reconfiguration}\\
    {\bf Given:} Graph $G=(V,E)$; independent sets $I_{\sf init},I_{\sf fin}$ of size $k$; integer $\ell$ given in binary; a partition $V=\sqcup_{i=1}^k P_i$ of the vertex set.\\
    {\bf Parameter:} $k$.\\
    {\bf Question:} Does there exist a sequence $I_{\sf init}=I_0,I_1,\dots,I_{T}=I_{\sf fin}$ of independent sets of size $k$ with $T\leq \ell$ and $|I_t\cap P_i|=1$ for all $t\in [0,T]$ and $i\in [1,k]$, such that for all $t\in [1,T]$, $I_{t}=I_{t-1}\setminus \{u\}\cup \{v\}$ for some $uv\in E(G)$ with $u\in I_{t-1}$ and $v \not\in I_{t-1}$? 
\end{verse}

\begin{theorem}
\label{thm:xnl_IS}
{\sc Binary Timed Partitioned TS-Independent Set Reconfiguration} is XNL-complete.
\end{theorem}

\begin{proof}

We first show that {\sc Binary Timed Partitioned TS-Independent Set Reconfiguration} is in XNL, that is, it can be modelled by a Nondeterministic Turing Machine using a work tape of size $\mathcal{O}(k \log n)$. One can store the current independent set of size $k$ on the work tape and allow only transitions between an independent set $I$ to an independent set $I'=I\setminus \{v\}\cup \{w\}$ if $vw\in E$, $v\in I$ and $w\not\in I$. We can generate the possible independent sets adjacent to a given independent set $I$ and keep track of the number of moves on a work tape of size $\mathcal{O}(k\log n)$. Since the number of independent sets of size $k$ is at most $n^k$, and a shortest sequence consists of distinct independent sets, we may assume that $\ell\leq n^k$.

To prove that {\sc Binary Timed Partitioned TS-Independent Set Reconfiguration} is XNL-hard, we give a reduction from \longchain.
Recall Theorem~\ref{thm:xnl_CNF} that states that \longchain is XNL-complete. 

Let $(q,r,F,P^1,\dots,P^k)$ be an instance of \longchain. We will create an instance $\Gamma$ of {\sc Binary Timed Partitioned TS-Independent Set Reconfiguration} with $3k+1$ token sets. The idea is to represent the choice of which variables $x_j^{(t)}$ are set to true with variable gadgets, and to create a clause checking gadget that verifies that  $F(x_1^{(t)},\dots,x_q^{(t)},x^{(t+1)}_1,\dots,x^{(t+1)}_q)$ is true. The time counter gadget has $k$ tokens, which together represent the integer $t$. Using the time constraint, we ensure that we have to follow a very specific sequence of moves, and can therefore not change which $x_j^{(t)}$ is true after we passed an independent set that made a choice for this. 

\subparagraph*{Time counter gadget.} We create $k$ \textit{time tokens} who have its token set within the time counter gadget, where the positions of these tokens represent an integer $t\in [1,r]$ with $r\leq q^k$. We create $k$ timers, consisting each of $4q$ vertices. For $i\in [1,k]$, the timer $t^i$ is a cycle on vertices  $t^i_0,\dots,t^i_{4q-1}$, which forms a token set for one of the time tokens. If the time tokens are on the vertices $t^1_{\ell_1},\dots, t^k_{\ell_k}$, then this represents the current time as 
$$
t=\sum_{i=1}^k (\ell_i \bmod q) q^{i-1}.
$$
Henceforth, we will silently assume $t$ to be given by the position of the time tokens as specified above. How these timers are connected such that they work as expected will be discussed later.

\subparagraph*{Variable gadget.}
We create four sets $A=\{a_1,\dots,a_q\}$, $B=\{b_1,\dots,b_q\}$, $C=\{c_1,\dots,c_q\}$ and $D=\{d_1,\dots,d_q\}$ that all contain $q$ vertices. These sets will be used to model which variables $x_j^{(t)}$ are chosen to be true.
We partition the sets in the same way as the variables, setting $A^i=\{a_j: j \in P^i\}$ for all $i\in[1,k]$ and defining  $B^i$, $C^i$ and $D^i$ similarly.

For all $i\in [1,k]$, we make $(A^i,B^i)$ and $(C^i,D^i)$ complete bipartite graphs, adding the edges $a_jb_{j'}$ and $c_jd_{j'}$ for all $j,j' \in P^i$. We specify $A^i\cup B^i $ and $C^i\cup D^i$ as token sets, and refer to the corresponding $2k$ tokens as \emph{variable tokens}. The first set is used to model the choice of the true variable $x_j^{(t)}$ for $j\in P^i$ for all odd $t$, whereas the second partition models the same for any even $t$. 

We will enforce the following. Whenever we check whether all the clauses are satisfied, we will either restrict all tokens of $A\cup B$ to be in $A$, or restrict all to be in $B$. Whenever we have to choose a new set of true variables for $t$ odd, we move all tokens from $A$ to $B$ (or the other way around). This movement takes exactly $k$ steps. The same holds for even $t$ and the sets $C$ and $D$.  

\subparagraph*{Clause checking gadget.} The clause checking gadget exists of four parts, called $AC$, $BC$, $BD$ and $AD$, named after which pair of sets we want the variable tokens to be in. All the vertices of the clause checking gadget form a token set, and we refer to the corresponding token as the \emph{clause token}. 
The token will traverse the gadget parts in the order $AC\to BC \to BD \to AD \to AC \to \dots$. 
If the token is on  $AC$, we require the variable tokens to be in $A$ and $C$ and we then check whether the clauses hold for the given choice of variables. 
The other parts are constructed likewise. For an example we refer to Figure~\ref{fig:XNLproofIS}.
\begin{figure}[h]
\centering
\scalebox{0.9}{
\begin{tikzpicture}[scale=.9,
dot/.style = {circle, draw, minimum size=#1,
              inner sep=0pt, outer sep=0pt},
dot/.default = .8 cm,
dot2/.style = {circle, draw, minimum size=#1,
              inner sep=0pt, outer sep=0pt},
dot2/.default = .5 cm]

    \node[dot2] (a1) at (0,10) {\small$a_1$};
    \node[dot2] (a2) at (0,9) {\small$a_2$};
    \node[dot2] (a3) at (0,8) {\small$a_3$};
    
    \node[dot2] (b1) at (1,10) {\small$b_1$};
    \node[dot2] (b2) at (1,9) {\small$b_2$};
    \node[dot2] (b3) at (1,8) {\small$b_3$};
    
    \node[dot2] (c1) at (4,10) {\small$c_1$};
    \node[dot2] (c2) at (4,9) {\small$c_2$};
    \node[dot2] (c3) at (4,8) {\small$c_3$};
    
    \node[dot2] (d1) at (5,10) {\small$d_1$};
    \node[dot2] (d2) at (5,9) {\small$d_2$};
    \node[dot2] (d3) at (5,8) {\small$d_3$};
    
    \draw[gray] (a1) to (b1);
    \draw[gray] (a1) to (b2);
    \draw[gray] (a1) to (b3);
    \draw[gray] (a2) to (b1);
    \draw[gray] (a2) to (b2);
    \draw[gray] (a2) to (b3);
    \draw[gray] (a3) to (b1);
    \draw[gray] (a3) to (b2);
    \draw[gray] (a3) to (b3);
    
    \draw[gray] (c1) to (d1);
    \draw[gray] (c1) to (d2);
    \draw[gray] (c1) to (d3);
    \draw[gray] (c2) to (d1);
    \draw[gray] (c2) to (d2);
    \draw[gray] (c2) to (d3);
    \draw[gray] (c3) to (d1);
    \draw[gray] (c3) to (d2);
    \draw[gray] (c3) to (d3);
    
    \node[dot] (T) at (-1,6) {\small$T^{AC}$};
    \node[dot2] (C1) at (2.5,7) {\small $v^1_1$};
    \node[dot2] (C2) at (2.5,6) {\small $v^1_2$};
    \node[dot] (T2) at (6,6) {\small$T^{BC}$};
    
    \draw[gray] (C1) to (T);
    \draw[gray] (C2) to (T);
    \draw[gray] (T2) to (C1);
    \draw[gray] (T2) to (C2);
    
    \draw[blue] (T).. controls (-1,11) .. (b1);
    \draw[blue] (T) .. controls (-1,10) .. (b2);
    \draw[blue] (T) .. controls (-1,9) .. (b3);
    
    \draw[blue] (T2).. controls (6,11) .. (d1);
    \draw[blue] (T2) .. controls (6,10) .. (d2);
    \draw[blue] (T2) .. controls (6,9) .. (d3);

    \draw[red] (C1).. controls (-.65,7.5) .. (a2);
    \draw[red] (C1) .. controls (0,7.6) .. (a3);
    \draw[red] (C1).. controls (2,9) .. (b2);
    \draw[red] (C1) .. controls (1,7.5).. (b3);
    \draw[red] (C1).. controls (2,10) .. (b1);
    \draw[red] (C1).. controls (3,11) .. (d1);
    \draw[red] (C1).. controls (3,10) .. (d2);
    \draw[red] (C1) .. controls (3,9) .. (d3);

    \draw[green!60!black] (C2).. controls (6,6.5) .. (d1);
    \draw[green!60!black] (C2) .. controls (5.8,6.5) .. (d2);
    \draw[green!60!black] (C2) .. controls (5.6,6.5) .. (d3);
    
    \draw[green!60!black] (C2).. controls (1.5,7) and (1.5,10) .. (b1);
    \draw[green!60!black] (C2) .. controls (1.5,7) and (1.5,9) .. (b2);
    \draw[green!60!black] (C2) .. controls (1.5,7) and (1.5,8) .. (b3);
    
    \draw[green!60!black] (C2).. controls (3.5,7) and (3.5,10) .. (c1);
    \draw[green!60!black] (C2) .. controls (3.5,7) and (3.5,9) .. (c2);
    
\end{tikzpicture}
}
\caption{Sketch of part of the construction of Theorem~\ref{thm:xnl_IS}. Given are the two variable gadgets for $A^i\cup B^i$ and $C^i \cup D^i$, the AC part of the clause checking gadget with one clause: $C_1=(x^{(t)}_1,x^{(t+1)}_3)$, where $t$ is odd. Hence $v^1_1$ checks whether $a_1$ is set to true and $v^1_2$ checks whether $c_3$ is set to true. }
\label{fig:XNLproofIS}
\end{figure}

We now give the construction of this gadget. We create a vertex, $T^{AC}$ that is connected to all $b\in B$. 
This ensures that if the clause token is on $T^{AC}$, all tokens from $A\cup B$ are on vertices in $A$, yet tokens will be able to move from vertices in $D$ to vertices in $C$.

Suppose $F= C_1 \land \dots \land C_S$ with each $C_i$ a disjunction of literals. Let $s\in [1,S]$ and let $C_s=y_1\lor \dots \lor y_{H_s}$ be the $s$th clause. 
We create a vertex $v^s_{h}$ for all $h\in[1,H_s]$. 
All $v^s_h$ are connected to all vertices in $B$ and $D$, which ensures that whenever the clause token is on some $v^s_h$, all variable tokens to be on vertices in $A$ and $C$ and prohibits these variable tokens to move. 

Let $h\in [1,H_s]$ and let $j\in [1,q]$ be such that $y_h$ is the $j$th variable.  We ensure that the clause token can only be on $v^s_h$ if the corresponding $x^{(t)}_j$ is modelled as true, that is, the corresponding variable token is on the vertex $a_j$ or $c_j$ (depending on the parity of $t$). To ensure this, we connect $v^s_h$ to all variables in $A^i\setminus\{a_j\}$ if $t$ is odd and to all variables in $C^i\setminus\{c_j\}$ if $t$ is even, where $i\in [1,k]$ satisfies $j\in P^i$.  

We add edges such that $(\{v_h^s\}_{h \in [1,H_s]},\{v_h^{s+1}\}_{h \in [1,H_{s+1}]})$ forms a complete bipartite graph for all $s\in[1,S-1]$. We connect $T^{AC}$ to all $v^1_{h}$ and we connect all $v^{S}_h$ to $T^{BC}$, the first vertex of the next gadget. 
Whenever we move the clause token from $T^{AC}$ to $T^{BC}$, we have to traverse a vertex $v^s_h$ for each clause $C_s$, which ensures that the literal $y_h$ in the clause $C_s$ is set to true according to the variable tokens. 

The gadget parts for $BC$, $BD$ and $AD$ are constructed likewise. We omit the details.

\subparagraph*{Connecting the time counter gadget.}
We now describe how to connect the vertices in the time counter gadget to those in the clause checking gadget. In the first timer, we create the following edges for $z\in [0,4q-1]$:
\begin{align*}
T^{AC} t^1_z &~~~\text{ when }z\equiv 2 \text{ or }z \equiv 3 \bmod 4,\\
T^{BC} t^1_z &~~~\text{ when }z\equiv 3 \text{ or }z \equiv 0 \bmod 4,\\
T^{BD} t^1_z &~~~\text{ when }z\equiv 0 \text{ or }z \equiv 1 \bmod 4,\\
T^{AD} t^1_z &~~~\text{ when }z\equiv 1 \text{ or }z \equiv 2 \bmod 4.
\end{align*}
This ensures that we can only move the first time token from $t_0^1$ to $t_{1}^1$ if the clause token is on $T^{AC}$, and that we cannot put the clause token on $T^{BC}$ before having moved the time token. To enforce that the time token moves when the clause token is at $T^{AC}$, we add edges between any $v^s_h$ vertex in this path and all $t^1_z$ with $z\not\equiv 1 \mod 4$. The edges are created in a similar manner for the paths following $T^{BC}$, $T^{BD}$ and $T^{AD}$. 

When the first time token has made $q$ steps, we allow the second time token to move $1$ step forward. 
For $i\in[2,k]$ we add all edges $t^i_zt^{i+1}_y$, \emph{except} for the following $y,z\in [0,4q-1]$:
\begin{align*}
 & y\equiv 0 \bmod 4 \text{ and }z\in[0,q],\\
  & y\equiv 1\bmod 4 \text{ and }z\in[q,2q],\\
  &y\equiv 2 \bmod 4 \text{ and }z\in[2q,3q] ,\\
 & y\equiv 3 \bmod 4 \text{ and } z\in[3q,4q-1]\cup\{0\}.
\end{align*}
This ensures, for example, that the $(i+1)$th  gadget token can move from $t^{i+1}_0$ to $t^{i+1}_{1}$ if and only if the $i$th time gadget token is on $t^{i}_q$. 

Finally, we add two sets $V_{\sf init}$ and $V_{\sf fin}$ of $2k$ vertices, and add the first set to the initial independent set $I_{\sf init}$ and the second to the final independent set $I_{\sf fin}$. Each vertex of $V_{\sf init}$ is added to the token set of $A^i\cup B^i$ or $C^i\cup D^i$ for some $i\in [1,2k]$, adding exactly one vertex to each token set, and similarly for $V_{\sf fin}$. 

We create edges $uv$ for all $u\in V_{\sf init}\cup V_{\sf fin}$ and $v$ in the clause checking gadget. 
We also create two vertices $c_{\sf init}$ and $c_{\sf fin}$ that are added to the initial and final independent set respectively, and to the token set of the clause token. We make $c_{\sf init}$ adjacent to $T^{AC}$ and $c_{\sf fin}$ to  a vertex $T^{XY}$, where $X,Y$ depend on the value of $r$ modulo $4$. 

The vertices  $c_{\sf init}$ and $c_{\sf fin}$ are adjacent to all vertices in the time gadgets except for those representing the time $0$ and $r$ respectively.
The initial independent set also contains the vertices in the time gadget that represent $t=0$ and similarly  $I_{\sf fin}$ contains the vertices that represent $r$.

\subparagraph*{Bounding the sequence length.} We give a bound $\ell$ on the length of the reconfiguration sequence, to ensure that only the required moves are made. Before moving the time token, we first move the $2k$ variable tokens into position. We can then move the clause token to $T^{AC}$, move the first time token so that the time represents $1$ and after that take $S+1$ steps to reach $T^{BC}$ (with $S$ the number of clauses in $F$), at which point we can move the first time token one step forward, and we need to move $k$ variable tokens from $A$ to $B$. Because we check exactly $r-1$ assignments, we need to move the $i$th time counter token exactly $\lfloor (r-1)/q^{k-(i-1)} \rfloor$ times. As a last set of moves, we need to move the variable tokens to the set $V_{\sf fin}$, and the clause token to $c_{\sf fin}$ taking another $2k+1$ steps. Hence, we set the maximum length of the sequence $\ell$ (from the input of our instance of {\sc Binary Timed Partitioned TS-Independent Set Reconfiguration}) to $$
4k+2+(r-1)(S+k+1) + \sum_{i=1}^k \lfloor (r-1)/q^{k-(i-1)} \rfloor.
$$
We claim that there is a satisfying assignment for our instance of \longchain~if and only if there is a reconfiguration sequence from $I_{\sf init}$ to $I_{\sf fin}$  of length at most $\ell$. It is not too difficult to see that a satisfying assignment leads to a reconfiguration sequence (by moving the variable tokens such that they represent the chosen true variables $x_j^{(t)}$ when the time tokens represent time $t$). 

Vice versa, suppose that there is a reconfiguration sequence of length $\ell$. 
This is only possible if the sequence 
takes a particular form: we need to move the time tokens for $ \sum_{i=1}^k \lfloor (r-1)/q^{k-(i-1)} \rfloor$ steps, and can only do this if we can move the clause token $(r-1)(S+1)+2$ steps. The moves of the clause token forces us to move $k$ variable tokens between $A$ and $B$ and between $C$ and $D$ a total of $(r-1)$ times, and we need a further $2k$ moves to get these from $V_{\sf init}$ and to $V_{\sf fin}$. In particular, there is no room for moving a variable token from one position in $A$ to another position in $A$, without the `time' having moved 4 places. Therefore, for each $i\in [1,k]$ and $t\in [1,r]$, there is a unique $j$ for which we find a variable token on $a_j\in A^i$, $b_j\in B^i$, $c_j\in C^i$ or $d_j\in D^i$ (which letter $a,b,c$ or $d$ we search for depends on the value of $t \text{ modulo } 4$) when the time tokens represent time $t$. 
This is the variable $x_j^{(t)}$ that we set to true from the $t$th variable set in partition $P^i$. 
\end{proof}

With Theorem~\ref{thm:equivalences}, we can now conclude

\begin{corollary}
{\sc Binary Timed TJ-Independent Set Reconfiguration} and {\sc Binary Timed TS-Independent Set Reconfiguration} are XNL-complete.
\end{corollary}

\subsection{Binary Timed Dominating Set Reconfiguration}
\label{section:ds-xnl}
\label{sec:XNLdom}

\begin{theorem}
{\sc Binary Timed Partitioned TS-Dominating Set Reconfiguration} is XNL-complete.
\end{theorem}
\begin{proof}
First, we note that the problem is in XNL, because it can be simulated by an Non-deterministic Turing Machine using $\mathcal{O}(k \log n)$ space. 

We prove the problem to be XNL-hard by giving a reduction from \longchain. The reduction is very similar to the reduction in Theorem~\ref{thm:xnl_IS}. Therefore, we will omit details and only give a short description.

Let $(q,r,F,P^1,\dots,P^k)$ be an instance of \longchain. We create an instance $\Gamma$ of {\sc Binary Timed TS-Dominating Set Reconfiguration} with $3k+2$ token sets. 

\subparagraph*{Variable gadget.} We create four set $A=\{a_1,\dots,a_q\}$, $B=\{b_1,\dots,b_q\}$, $C=\{c_1,\dots,c_q\}$ and $D=\{d_1,\dots,d_1\}$ that all contain $q$ vertices. We partition the sets in the same way as the variables, setting $A^i=\{a_j: j\in P^i\}$ for all $i\in [1,k]$ and defining $B^i$, $C^i$ and $D^i$ similarly. 

For all $i\in[1,k]$ we make $(A^i,B^i)$ and $(C^i,D^i)$ complete bipartite graphs and let $A^i \cup B^i$ and $C^i\cup D^i$ be token sets. We refer to these $2k$ tokens as the \emph{variable tokens}. 

\subparagraph*{Clause checking gadget}
The clause checking gadget exists of four parts, called $AC$, $BC$, $BD$ and $AD$. All the vertices of the clause checking gadget form a token set, and we refer to the corresponding token as the \emph{clause token}. We give a construction of the AC part, the other parts can be constructed likewise.

We create two vertices: $T^{AC}_1$ and $T^{AC}_2$, connected by an edge. $T^{AC}_1$ will allow the variable tokens to move, and $T^{AC}_2$ checks if the variable gadget tokens are on vertices in $A$ and $C$. To accomplish this, we connect $T^{AC}_1$ to \emph{all} vertices in the variable gadgets, allowing movement of the tokens. With $T^{AC}_2$ we check if the tokens are only on vertices in $A$ and $C$, by adding edges $T^{AC}_2a_j$ and $T^{AC}_2c_j$ for all $j\in[1,q]$. The vertices in $B$ and $D$ must then be dominated by vertices in $A$ and $C$ respectively.

Suppose $F = C_1 \wedge \cdots \wedge C_s$ with each $C_i$ a disjunction of positive literals. Let $s \in [1,S]$ and let $C_s = y_1 \vee \cdots \vee y_{H_s}$ be the $s$th clause. 
We create a vertex $v^s_h$ for all $h\in[1,H_s]$. 

Recall that $T^{AC}_2$ ensured that all tokens are on vertices in $A$ and $C$. Let $h \in [1,H_s]$ and let $j\in[1,q]$ be such that $y_h$ is the $j$th variable. We ensure that the clause token can only be on $v^s_h$ if the corresponding $x_j^{(t)}$ is modelled as true, by connecting $v^s_h$ to all $a_{j'}$ for $j'\in[1,q]\setminus\{j\}$ such that $a_j$ must be in the dominating set whenever $v^h_s$ is (for $t$ odd). Also, we connect $v^s_h$ to all $c \in C$, so that they are dominated. 

We add edges such that $(\{v^s_h\}_{h\in[1,H_s]},\{v^{s+1}_h\}_{h\in[1,H_{s+1}]})$ forms a complete bipartite graph for all $s \in [1,S-1]$. We connect $T_2^{AC}$ to all $v^1_h$ and we connect all $v^S_h$ to $T^{BC}_1$. 

\subparagraph*{Time counter gadget.}
We create a time counter gadget, keeping track of the integer $t$. For each timer $i\in[1,k]$ we add $4q+4$ vertices. From these, the vertices $t^i_0,\dots,t^i_{4q-1}$ form a cycle and are a token set for a token. We name the four additional vertices $\gamma^i_1$, $\gamma^i_2$, $\gamma^i_3$ and $\gamma^i_4$ and these will be the bridge between two consecutive timers $i-1$ and $i$.
We add edges $t^{i}_z\gamma^i_y$ for all $z\in[0,4q-1]$ and $y \in \{1,2,3,4\}$ such that:
\begin{align*}
    & z \equiv 0  \bmod 4 \text{ and } y \in \{1,2\},\\
    & z \equiv 1  \bmod 4 \text{ and } y \in \{2,3\},\\
    & z \equiv 2  \bmod 4 \text{ and } y \in \{3,4\},\\
    & z \equiv 3  \bmod 4 \text{ and } y \in \{1,4\}.
\end{align*}
Furthermore, we add edges $t^{i-1}_z\gamma^i_y$ for all $z\in[0,4q-1]$ and $y \in \{1,2,3,4\}$ \emph{except} for the following
\begin{align*}
    &y= 1 \text{ and } z \in [0,q-1],\\
    &y= 2 \text{ and } z \in [q,2q-1],\\
    &y= 3 \text{ and } z \in [2q,3q-1],\\
    &y= 4 \text{ and } z \in [3q,4q-1].
\end{align*}
For $i=1$, we add all edges between $T^{AC}_1$, $T^{BC}_1$, $T^{BD}_1$ and $T^{AD}_1$, and all $\gamma^1_y$ for $y\in[1,4]$ \emph{except} for  the edges $T^{AC}_1\gamma^1_1$, $T^{BC}_1\gamma^1_2$, $T^{BD}_1\gamma^1_3$ and $T^{AD}_1\gamma^1_4$. This construction has the same effect as the time counter gadgets constructed in the proof of Theorem~\ref{thm:xnl_IS}. 

We add two sets $V_{\sf init}$ and $V_{\sf fin}$ of $2k$ vertices and these to $D_{\sf init}$ and $D_{\sf fin}$ respectively. Each vertex of $V_{\sf init}$ is added to the token set of $A^i\cup B^i$ or $C^i\cup D^i$ for some $i\in [1,2k]$. Similarly for $V_{\sf fin}$. We create edges $uv$ for all $u \in V_{\sf init}\cup V_{\sf fin}$ and $v$ in the clause checking gadget. We also create two vertices $c_{\sf init}$ and $c_{\sf fin}$ that are added to $D_{\sf init}$ and $D_{\sf fin}$ respectively and to the token set of the clause token. We make $c_{\sf init}$ adjacent to $T_1^{AC}$ and $c_{\sf fin}$ to $T^{XY}_1$, where $X,Y$ depend on the value of $r \text{ modulo } 4$. 

The vertex $c_{\sf init}$ is adjacent to $\gamma^1_1$ and the vertex $c_{\sf fin}$ is adjacent to $\gamma^1_y$ where $y \equiv r \text{ modulo } 4$. The initial dominating set contains the vertices in the time gadget that represent $t=0$ and similarly $D_{\sf fin}$ contains the vertices that represent $r$.

\subparagraph*{Dominating vertex.}
The following addition does two things at the same time: ensuring the clause checking gadget vertices and timer cycle vertices are dominated and putting the $\gamma$ vertices of the time counter gadget in a token set. For this we add one new token set, consisting of all $\gamma$ vertices and three vertices $z_{\sf dom}$, $z_{\sf gar}$ and $z_{\sf gar}'$. The vertices $z_{\sf gar}$ and $z_{\sf gar}'$ are only connected to $z_{\sf dom}$, implying that $z_{\sf dom}$ is in the dominating set at all times. We then add edges between $z_{\sf dom}$ and all counter gadget tokens and all $t^i_j$ for $j\in [0,4q-1]$, to dominate them. We add $z_{\sf dom}$ to both $D_{\sf init}$ and $D_{\sf fin}$.

\subparagraph*{Bounding the sequence length.}
We set $\ell$, the maximum length of the sequence, to 
\[
4k +2 + (r-1)(S+k+2) + \sum_{i=1}^k \lfloor (r-1)/q^{k-(i-1)}\rfloor.\]
The analysis of this integer is almost identical to that of in Theorem~\ref{thm:xnl_IS}, except that one extra step per time step is required to move from $T^{XY}_1$ to $T^{XY}_2$.
\end{proof}

Again, with help of Theorem~\ref{thm:equivalences}, we can conclude the completeness for variants with other token move rules.

\begin{corollary}
{\sc Binary Timed TS-Dominating Set Reconfiguration} and {\sc Binary Timed TJ-Dominating Set Reconfiguration} are XNL-complete.
\end{corollary}

\section{XNLP-completeness of Timed Reconfiguration}
\label{subsection:reconfiguration}
In this section, we show that the considered reconfiguration problems are XNLP-complete, when the number of moves is
given in unary.

\subsection{Timed Independent Set Reconfiguration}
\label{appendix:reconfiguration}
\label{section:is-xnlp}

Instead of directly looking at reconfiguration of independent sets, we
present first an XNLP-completeness proof for {\sc Timed TJ-Clique Reconfiguration}, as the idea of the 
proof is most naturally phrased in terms of clique reconfiguration. After that, we deduce the XNLP-completeness for
the other variants from it.
The starting point for the reduction in this section is the following problem.

\begin{verse}
  {\sc Chained Multicolored Clique}\\
  {\bf Input:} Graph $G=(V,E)$; partition of $V$ into sets $V_1, \ldots, V_r$, such that
  for each edge $\{v,w\}\in E$, if $v\in V_i$ and $w\in V_j$, then $|i-j|\leq 1$; 
  function $f: V \rightarrow \{1,2,\ldots, k\}$.\\
  {\bf Parameter:} $k$.\\
  {\bf Question:} Is there a subset $W\subseteq V$ such that for each $i\in [1,r]$,
  $W \cap (V_i \cup V_{i+1})$ is a clique, and for each $i\in [1,r]$ and each $j\in [1,k]$,
  there is a vertex $w\in V_i\cap W$ with $f(w)=j$?
\end{verse}

\begin{theorem}[Bodlaender et al.~\cite{BodlaenderGNS21}]
{\sc Chained Multicolored Clique} is XNLP-complete.
\end{theorem}

\begin{theorem}
{\sc Timed TJ-Clique Reconfiguration} is XNLP-complete, with the number of steps given in unary.
\label{theorem:cliquereconfiguration}
\end{theorem}

\begin{proof}
Membership in XNLP is easy: one can guess the sequence of moves, and keep the current positions of the
tokens in memory.

For the hardness, we transform from {\sc Chained Multicolored Clique}.
Suppose that we are given an integer $k$ and a graph $G=(V,E)$
with a partition of $V$ into sets $V_1, \ldots , V_r$. 
Let $f: V \rightarrow [1,k]$ be a given vertex coloring.
We look for a chained multicolored clique $W \subseteq V$, that is,  for each $i\in [r]$, $|W\cap V_i|=k$, for each $i\in [1,r-1]$,
$W \cap (V_i \cup V_{i+1})$ is a clique, and for each $i\in [1,r]$ and $j\in [1,k]$, $W\cap V_i$ contains exactly one vertex
of color $j$.

We may assume that for each edge $\{v,w\}\in E$, if $v\in V_i$ and $w\in V_{i'}$, then $|i-i'|\leq 1$
and if $i=i'$, then $f(v)\neq f(w)$. (Edges that do not fulfill these properties will never contribute to a chained multicolored clique, and thus we can remove them. A transducer exists that removes all edges 
that do not fulfill the property using logarithmic space.)

We build a graph $H$ as follows.
\begin{itemize}
    \item Start with $G$.
    \item Add vertex sets $V_{-1}, V_0, V_{r+1}, V_{r+2}$, each of size $k$. Extend $f$ as follows: for each color $j\in [1,k]$, and each of the sets $V_{-1}, V_0, V_{r+1}, V_{r+2}$, take one vertex of the set and color it with $j$.
    \item For all $v\in V_i$ and $w\in V_{i'}$, $v\neq w$, add an edge only if at least one of the following holds: 
    \begin{itemize}
        \item $i \in \{-1,0\}$ and $i' \in \{-1,0\}$, i.e. $V_{-1}\cup V_0$ is a clique,
        \item $i = 0$ and $i' =1$, 
        \item $i \in \{r+1,r+2\}$ and $i' \in \{r+1,r+2\}$, i.e. $V_{r+1}\cup V_{r+2}$ is a clique,
        \item $i = r$ and $i' = r+1$, 
        \item $|i-i'|=2$, and $f(v)\neq f(w)$: a vertex is adjacent to all vertices `two sets away', except those with the same color.
    \end{itemize}
\end{itemize}

We reconfigure a clique with $2k$ vertices. The initial
configuration is $V_{-1} \cup V_0$ and the final configuration is $V_{r+1} \cup V_{r+2}$.

\begin{claim}
Suppose that $G$ has a chained multicolored clique with $k$ colors. Then we can reconfigure $V_{-1} \cup V_0$
to $V_{r+1} \cup V_{r+2}$ with $k \cdot (r+2)$ jumping moves in $H$, with each intermediate configuration a clique.
\end{claim}

\begin{proof}
Let $W$ be the chained multicolored clique with $k$ colors.
Write $W' = W \cup V_{-1} \cup V_0 \cup V_{r+1} \cup V_{r+2}$. 
The starting configuration is $S= V_{-1} \cup V_{0}$.

Take the following sequence:
for $i$ from $-1$ to $r$, for $j$ from $1$ to $k$, let $v$ be the vertex in $W' \cap V_i$ with
$f(v)=j$, and $w$ be the vertex in $W'\cap V_{i+2}$ with $f(v)=j$ and update $S$ to $S \cup \{w\} \setminus \{v\}$.

It is easy to verify that at each step $S$ is a clique; the total number of moves equals $k \cdot (r+2)$.
\end{proof}

\begin{claim}
Suppose that we can reconfigure $ V_{-1} \cup V_0$
to $V_{r+1} \cup V_{r+2}$ with $k \cdot (r+2)$ jumping moves in $H$, with each intermediate configuration a clique. Then $G$ has a chained multicolored clique with $k$ colors.
\end{claim}

\begin{proof}
The \emph{level} of a vertex $v\in V(G)$ is the unique $i\in [1,r]$ for which $v \in V_i$. Note that for each edge $\{v,w\}$ in $H$, if $v\in V_i$ and $w\in V_{i'}$, then $|i-i'|\leq 2$. Thus, each clique can contain vertices of at most three consecutive levels. Also, vertices in the same level with the same color are not adjacent, and vertices two levels apart with the same color are not adjacent.
Thus, if $H$ has a clique with $2k$ vertices, then this clique is a subset of $V_i \cup V_{i+1} \cup V_{i+2}$ for some $i\in [-1,r]$, with the property that for each color $j\in [1,k]$, there is one vertex with this
color in $V_i \cup V_{i+2}$ and one vertex with this color in $V_{i+1}$.

Denote the initial, final and intermediate clique configurations by $S$. To each such set $S$, we associate a value. The \emph{potential} $\Phi(S)$ is the sum of all levels of the vertices in a clique $S$. 

We now argue that each move of a token can increase $\Phi(S)$ by at most two. For any clique $S$, there is an $i$ with $S\subseteq V_i \cup V_{i+1} \cup V_{i+2}$ and $S\cap V_i\neq \emptyset$. If $|S\cap V_i|\geq 2$, then after a move of one vertex, there still will be at
least one vertex in $S\cap V_i$. This means that we can only move to vertices in $V_{i'}$ with
$i'\leq i+2$. If $|S\cap V_i| = 1$, then suppose $S\cap V_i=\{v\}$ with $f(v)=j$. Now, if we move
$v$ then we can only move $v$ to a vertex in $V_i \cup V_{i+2}$ with color $j$. In both cases, the potential of $S$ increases by at most two.

The initial configuration $S_0$ has $\Phi(S_0) = -k$ and the final configuration $S_{k(r+2)}$ has
$\Phi(S) = k \cdot (2r+3)$. Thus, each of the $k \cdot (r+2)$ moves must increase the potential by
exactly two. Thus, each move must take a vertex and move it to a vertex exactly two levels later.
As each clique configuration spans at most three levels, we have that after each $k$th move, the configuration spans two levels, and necessarily is a clique with for each color, one vertex from each of these two levels. The collection of these cliques, except the first and last two, forms a chained multicolored clique in $G$. 
\end{proof}
To finish the hardness proof, we combine the two claims above and observe that $H$ can be constructed in logarithmic space.
\end{proof}

\begin{corollary}
{\sc Timed TJ-Independent Set Reconfiguration} and {\sc Timed TS-Independent Set Reconfiguration} are XNLP-complete.
\end{corollary}

\begin{proof}
Taking the complement of a graph can be done in logarithmic space; thus, the result follows directly
from Theorem~\ref{theorem:cliquereconfiguration}, noting that we only `jump' over non-edges in the proof of Theorem \ref{theorem:cliquereconfiguration}, so that in the complement the reconfiguration sequence adheres to the token sliding rules as well. (Instead of the last argument, we can also use Theorem~\ref{thm:equivalences}.)
\end{proof}

For the clique variant, a slight alteration to the current construction works for the token sliding rule.

\begin{corollary}
{\sc Timed TS-Clique Reconfiguration} is XNLP-complete.
\end{corollary}
\begin{proof}[Sketch]
We follow the proof of Theorem \ref{theorem:cliquereconfiguration}. We build the graph $H$ in a similar fashion, apart from the edges between pairs of sets $(V_{i},V_{i+2})$. For each $i\in [-1,r]$, we place an edge from $v\in V_{i}$ to $w\in V_{i+2}$ if and only if  $f(v)\geq f(w)$. 

Inductively, one can show that any clique must still span at most three consecutive layers and that each clique contains exactly one vertex of each color from the even and odd levels at any point. Indeed, the vertex of color $k$ in $V_{-1}$ can only slide to a vertex of color $k$ in $V_1$, and since the vertices of color $k$ in $V_1$ form an independent set, this forces the vertex of color $k-1$ in $V_{-1}$ to slide to a vertex of color $k-1$ in $V_1$ etcetera. 

The same argument can be used for turning a clique into a configuration sequence;
however, now it has become important that we slide the vertices in the right order, that is, the first token to be moved from $V_i$ to $V_{i+2}$ needs to be the one on a vertex of color $1$ (to ensure we remain a clique at any point). 
\end{proof}

\subsection{ Timed Dominating Set Reconfiguration}
\label{section:ds-xnlp}

As starting point for the reduction that shows {\sc Timed TS-Dominating Set Reconfiguration} to be XNLP-complete,
we use the following problem. 
\begin{verse}
{\sc Partitioned Regular Chained Weighted Positive CNF-Satisfiability}\\
{\bf Input}: $r$ sets of Boolean variables $X_1, X_2, \ldots X_r$, each of size $q$; an integer $k\in \N$; Boolean formula $F_1$, which is in conjunctive normal form with only positive literals, and an expression on $2q$ variables; for each $i$,
a partition of $X_i$ into $X_{i,1}, \ldots, X_{i,k}$ with for all $i_1,i_2$, $j$: $|X_{i_1,j}| = |X_{i_2,j}|$.\\
{\bf Parameter}: $k$.\\
{\bf Question}: Is it possible to satisfy the formula 
\[ \bigwedge_{1\leq i\leq r-1} F_1(X_i,X_{i+1}) \]
by setting from each set $X_{i,j}$ exactly $1$ variable to true and all others to false?
\end{verse} 

\begin{theorem}[Bodlaender et al.~\cite{BodlaenderGNS21}]
{\sc Partitioned Regular Chained Weighted Positive CNF-Satisfiability} is
XNLP-complete.
\end{theorem}

\begin{theorem}
\label{thm:DSreconfigTS}
{\sc Timed TS-Dominating Set Reconfiguration} is XNLP-complete.
\end{theorem}
\begin{proof}
The problem is in XNLP follows since we can `explore' all dominating sets that can be reached from $S$ within $T$ steps non-deterministically, storing the current positions of the $k$ counters and the number of steps taken so far using $O(k\log n)$ bits, accepting if the current position encodes $S'$ and halting after $T$ steps. 

We prove hardness via a reduction from {\sc Partitioned Regular Chained Weighted Positive CNF-Satisfiability}.
We obtain $r$ sets of Boolean variables $X_1, \ldots, X_r$, a Boolean formula $F_1$, and a partition of
each set $X_i$ into sets $X_{i,1},\dots,X_{i,k}$, each of size $q$. We will assume that $r$ is even. We will create a graph and dominating sets $S,S'$ of size $2k+2$ such that $S$ can be reconfigured to $S'$ in $T=\frac{5}2r-2$ steps if and only if $\wedge_{1\leq i\leq r-1} F_1(X_i,X_{i+1})$ can be satisfied
by setting exactly one variable to true from each set $X_{i,j}$.

The idea of the construction is to create a `timer' of $2r-2$ steps, of which each dominating set in the sequence must contain exactly one element, and enforce that the first dominating set $S$ contains the start vertex $t_0$ whereas the second dominating set $S'$ contains the ending vertex $t_{2r-3}$. This means the token on $t_0$ needs to slide over the timeline, and we use the time constraint to enforce this can happen in a single fashion, using the vertices $t_{2i}$ to `move $k$ tokens from $X_{i}$ to $X_{i+2}$' and the vertex $t_{2i-1}$ to `verify whether the formula $F_1(X_{i},X_{i+1})$ holds'.

\begin{figure}
    \centering
    \includegraphics{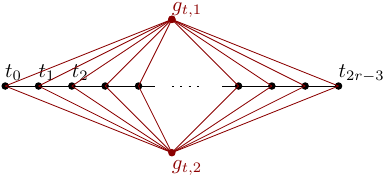}
    \caption{In the time line, at least one vertex needs to get chosen in order to dominate the time guardian vertices $g_{t,1}$ and $g_{t,2}$.}
    \label{fig:dsreconf1}
\end{figure}
Our construction has the following parts:
\begin{itemize}
    \item A `timer': a path with vertices $t_0, t_1, \ldots, t_{2r-3}$, along with two time guardians $g_{t,1}, g_{t,2}$ that are adjacent to $t_i$ for all $i\in [0,2r-3]$ and nothing else. In order to dominate the two time guardians using only a single vertex, we need to choose one of the $t_i$. This part of the construction is illustrated in Figure \ref{fig:dsreconf1}.
    \item For each variable $x$ a vertex $v_x$.
    \item A `dominator' vertex $d$ with a pendant vertex $p_d$ (which is only adjacent to $d$ and enforces that we must always use one token for $d$ or $p_d$). We use $d$ to dominate all the $t_i$ and all the variable vertices.
    \item For each clause $\phi$ in the Boolean formula $F_1$, a vertex $w_{\phi,i}$ for all $i\in [1,r-1]$. The vertex $w_{\phi,i}$ is adjacent to vertex $v_x$ for $x\in (X_i\cup X_{i+1})\cap \phi$ (that is, setting $x$ to true satisfies $\phi$). Moreover, $w_{\phi,i}$ is adjacent to all $t_j$ with $j\neq 2i-1$.
    \item Vertices $m_{i,j}$ for $i\in [1,r], j\in [1,k]$. These will ensure that we move tokens from variables in $X_i$ to variables in $X_{i+2}$ at certain time steps.
    \item Vertices $g_{oe,0,j,a}$ and $g_{oe,1,j,a}$ for all $j\in [1,k]$ and $a\in \{1,2\}$. For $a\in \{1,2\}$ and $j\in [1,k]$, the vertex $g_{oe,0,j,a}$ (respectively $g_{oe,1,j,a}$) is adjacent to all vertices $v_x$ for $x\in X_{i,j}$ with $i$ even (respectively odd) and nothing else; these ensure that for each `variable group' we always have at least one token on a vertex corresponding to this group.
    \item Start vertices $s_1,\dots,s_{2k}$ to add to the initial dominating set $S$. The vertices $s_1,\dots,s_k$ are adjacent to $v_x$ for all $x\in X_1$. The vertices $s_{k+1},\dots,s_{2k}$ are adjacent to $v_x$ for all $x\in X_2$.
    \item End vertices $s_1',\dots,s_{2k}'$ to add to the final dominating set $S'$.
    The vertices $s_1',\dots,s_k'$ are adjacent to $v_x$ for all $x\in X_{r-1}$. The vertices $s_{k+1}',\dots,s_{2k}'$ are adjacent to $v_x$ for all $x\in X_r$.
\end{itemize}
There are some further edges as explained later.
An outline of the construction is depicted in Figure \ref{fig:dsreconf2}.

\begin{figure}
    \centering
    \includegraphics{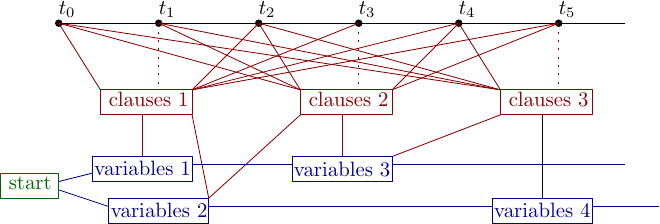}
    \caption{An overview of the construction is given.}
    \label{fig:dsreconf2}
\end{figure}
Due to the dominator vertex and its pendant vertex, we do not have to worry about dominating the vertices $t_i$ and the variable vertices. 

The guardian vertices enforce that we need to place a minimum number of tokens on certain vertex sets in order to have a dominating set. One token is always on the dominator vertex; we will call this the \emph{dominator token}. We need $2k$ tokens on various variable sets, which we refer to as \emph{variable tokens} and one token on one of the $t_i$, which we refer to as the \emph{time token}. Since we only have $2k+2$ tokens, there are no further tokens available and we need to place exactly these tokens at each time step. In particular, at each time step we can either move the timer token within the timer, or move the variable tokens within their respective sets.

We set $S=\{t_0,d,s_1,\dots,s_{2k}\}$ and set $S'=\{t_{2r-3},d,s_1',\dots,s_{2k}'\}$.

For $i\in [1,r-1]$, the clause vertices $w_{\phi,i}$ are non-adjacent to $t_j$ if $j\neq 2i-1$; this ensures that when the time token is at $t_{2i-1}$, the variable tokens need to take care of dominating $w_{\phi,i}$ (which happens if and only if they are placed in a satisfying assignment).

For $i\in [1,r-2]$, we make $v_x$ adjacent to $v_y$ for all $x\in X_i$ and $y\in X_{i+2}$. This allows us to `slide' variable tokens from $X_1$ to $X_3$ to $X_5$ etcetera. For $i\in [3,r]$, we ensure that variable tokens are moved from $X_{i-2}$ to $X_{i}$ when the time token is at $t_{2(i-2)}$ using the move vertices $m_{i,\ell}$; these are adjacent to $t_j$ for all $j\neq 2(i-2)+1$ and to all $v_x$ for $x\in X_{i,\ell}$. For $i\in [1,2]$, the move vertices $m_{i,\ell}$ are adjacent to $v_x$ for all $x\in X_{i,\ell}$ and to all $t_j$ for $j\neq 1$. Before the time token can be moved from $t_0$ to $t_1$, the tokens on the start vertices need to be moved from $S$ to the $2k$ variable sets $X_{i,\ell}$ for $i\in \{1,2\},\ell\in [1,k]$ in order to dominate the $m_{i,\ell}$. 

To move tokens from $S$ to $S'$, we need to move the $2k$ tokens via the variable vertices. This takes at least $r/2+1$ moves, as each token can go through either the odd or the even variable sets. The token on $t_0$ has to move to $t_{2r-3}$, which takes at least $2r-3$ moves. 
We set the time bound to $T=r/2+1+2r-3$ so that we are enforced to exactly take these steps. In particular, for each $i\in [1,r]$ and $\ell\in [1,k]$, there is a unique $x\in X_{i,\ell}$ for which a variable token gets placed on $v_x$ for some $x\in X_{i,\ell}$ in a valid reconfiguration sequence.

In conclusion, there is a reconfiguration sequence from $S$ to $S'$ within $T$ steps if and only if we can select one variable per $X_{i,\ell}$ so that when these are set to true, the formula $\wedge_{1\leq i\leq r-1} F_1(X_i,X_{i+1})$ is satisfied.
\end{proof}





Theorem~\ref{thm:equivalences} now also gives the following result.

\begin{corollary}
{\sc Timed TJ-Dominating Set Reconfiguration} is XNLP-complete.
\end{corollary}

\section{Membership in $W[1]$ and $W[2]$ for Combined Parameterizations}
\label{section:membershipw1w2}

\subsection{Independent Set with Combined Parameterization}
\label{section:is-w}
In this section, we give a proof for membership in $W[1]$ for
{\sc Timed TJ-Independent Set Reconfiguration} with the number of tokens and length of the reconfiguration sequence as combined parameter. We do this by showing that it can be formulated as an instance of {\sc Weighted $3$-CNF-Satisfiability}. Downey
and Fellows showed that {\sc Weighted $3$-CNF-Satisfiability} is $\mathrm{W}[1]$-complete \cite{DowneyF99}.

\begin{verse}
{\sc Weighted $3$-CNF-Satisfiability}\\
{\bf Given:} Boolean formula $F$ on $n$ variables in conjunctive normal form such that each clause contains at most $3$ literals; integer $K$.\\
{\bf Parameter:} $K$.\\
{\bf Question:} Can we satisfy $F$ by setting exactly $K$ variables to true?
\end{verse}

\begin{lemma}
{\sc Timed TJ-Independent Set Reconfiguration}  with the number of tokens and length of the reconfiguration sequence as combined parameter is in $W[1]$.
\end{lemma}

\begin{proof}
Let $(G= (V,E), I_{\sf init}, I_{\sf fin}, k, \ell)$ be an instance of {\sc Timed TJ-Independent Set Reconfiguration}.
We set $C=(k+1+\ell)^2$ and $K=\ell (C+1)+(\ell+1)k$.
We add the following variables to an {\sc Weighted 3-CNF-Satisfiability} instance for all $t\in [0,\ell]$:
\begin{itemize}
    \item $s_{t,v}$, for each vertex $v\in V$. This should be set to true if and only if $v$ has a token at time $t$.
    \item $m_{t,v,w}^{(i)}$, for each pair of distinct vertices $v,w\in V$ and for all $i\in [1,C]$. This should be set to true if and only if we move a token from $v$ to $w$ from time $t-1$ to time $t$.
    \item $m_{t,\emptyset}^{(i)}$, for all $i\in [1,C]$. This is set to true if no token is moved at from time $t-1$ to time $t$. 
    \item $a_{t,v}$ for all $v\in V$. This is set to true if and only if $v$ received a token from time $t-1$ to time $t$.
    \item $a_{t,\emptyset}$. This is set to true if no vertex received a token from time $t-1$ to time $t$.  
\end{itemize}
We add clauses that are satisfied if and only if the set of true variables corresponds to a correct TJ-reconfiguration sequence from $I_{\sf init}$ to $I_{\sf fin}$. 
\begin{itemize}
    \item We have clauses with one literal that ensure that at time 0, we have the initial configuration: for each $v\in I_{\sf init}$, we have a clause $s_{0,v}$ and for each $v\not\in I_{\sf init}$, we have a clause $\neg s_{0,v}$.
    \item Similarly, we have clauses that ensure that at time $\ell$, we have the final configuration:
    for each $v\in I_{\sf fin}$, we have a clause $s_{\ell,v}$ and for each $v\not\in I_{\sf fin}$, we have a clause $\neg s_{\ell,v}$.
    \item All $m_{t,\star}^{(i)}$ are equivalent: for all distinct $i,j\in [1,C]$, for all
    $t\in [1,\ell]$ and for all distinct $v,w \in V$, we add the clauses
    $\neg m_{t,v,w}^{(i)}\vee m_{t,v,w}^{(j)}$ and $ m_{t,v,w}^{(i)}\vee \neg m_{t,v,w}^{(j)}$.
    For all distinct $i,j \in [1,C]$, for all $t\in [1,\ell]$, we add the clauses
    $\neg m_{t,\emptyset}^{(i)}\vee m_{t,\emptyset}^{(j)}$ and $ m_{t,\emptyset}^{(i)}\vee \neg m_{t,\emptyset}^{(j)}$.
    \item We have clauses that ensure that at each time $t\in [1,\ell]$, at most one move is selected: for any two distinct pairs of distinct vertices $(v,w)$ and $(v',w')$, we add the clauses $\neg m_{t,v,w}^{(1)} \vee \neg m_{t,v',w'}^{(1)}$ and $\neg m_{t,v,w}^{(1)} \vee \neg m_{t,\emptyset}^{(1)}$. 
    \item For $t\in [1,\ell]$, if the move $m_{t,v,w}^{(1)}$ is selected, then $v$ lost a token and $w$ obtained a token from time $t-1$ to time $t$: $\neg m_{t,v,w}^{(1)}\vee s_{t-1,v}$, $\neg m_{t,v,w}^{(1)}\vee \neg s_{t-1,w}$, $\neg m_{t,v,w}^{(1)}\vee \neg s_{t,v}$ and $\neg m_{t,v,w}^{(1)}\vee  s_{t,w}$.
    \item For $t\in [1,\ell]$, tokens on vertices not involved in the move remain in place. For all distinct $v,w,u\in V$, we add the clauses
    \begin{align*}
        &\neg m_{t,\emptyset}^{(1)} \vee \neg s_{t-1,v} \vee s_{t,v},\\
        &\neg m_{t,\emptyset}^{(1)} \vee  s_{t-1,v} \vee \neg s_{t,v},\\
        &\neg m_{t,v,w}^{(1)} \vee \neg s_{t-1,u} \vee s_{t,u} \text{ and }\\
        &\neg m_{t,v,w}^{(1)} \vee s_{t-1,u} \vee \neg  s_{t,u}.
    \end{align*}
    \item We record if a token was added to a vertex: for all $t\in [1,\ell]$ and $v\in V$, we add the clause $s_{t-1,v}\vee \neg s_{t,v} \vee a_{t,v}$. This in particular ensures that $a_{t,v}$ is true when $m_{t,v,w}^{(1)}$ is true for some vertex $w\neq v$.
    \item No move happened if and only if no token was added: for all $t\in[1,\ell]$ we add the clauses $\neg m^{(1)}_{t,\emptyset} \vee a_{t,\emptyset}$ and $\neg  a_{t,\emptyset} \vee m^{(1)}_{t,\emptyset}$.
    \item At most one $a_{t,\star}$ is set to true, implying that at most one taken gets added at each time step:  for all $t\in [1,\ell]$ and distinct $v,w\in V$, we add the clauses $\neg a_{t,v}\vee \neg a_{t,w}$ and $\neg a_{t,v} \vee \neg a_{t,\emptyset}$.
    \item Finally, we check whether the current set forms an independent set: for all edges $vw\in E(G)$ and $t\in [0,\ell]$, we add the clause $\neg s_{t,v} \vee \neg s_{t,w}$.
\end{itemize}

If there is a TJ-independent set reconfiguration sequence $I_{\sf init}=I_0,\dots,I_T=I_{\sf fin}$ with $T\leq \ell$, then  we set $s_{t,v}$ to true if and only if $v\in I_t$ for $t\in  [0,T]$. For all $t\in [T,\ell]$, we set $s_{t,v}$ to true if and only if $v\in I_{T}$. 

Let $t\in [1,\ell]$. If $I_t=I_{t-1}$, we set $a_{t,\emptyset}$ to true and $m^{(i)}_{t,\emptyset}$ to true for all $i\in [1,C]$. Otherwise, we find $I_{t}=I_{t-1}\setminus \{v\}\cup \{w\}$ for some $v,w\in V$ and we set $m_{t,v,w}^{(i)}$ and $a_{t,v}$ to true for all $i\in [1,C]$. All other $m_{t,\star}^{(i)}$ are set to false. This gives a satisfying assignment with exactly $\ell (C+1)+(\ell+1)k=K$ variables set to true.

Suppose now that there is a satisfying assignment with $K$ variables set to true. At most one $a_{t,v}$ variable can be true for each $t\in [1,\ell]$.
Exactly $k$ variables of the form $s_{0,v}$ are set to true by the initial condition. If there are $k'$ tokens true at time $t$, then there are at most $k'+1$ tokens true at time $t+1$ and so the $s_{t,v}$ and $a_{t,v}$ variables together can constitute at most $((k+\ell)+1)\ell\leq C-1$ true variables. Therefore, there must be strictly more than $(C+1)(\ell-1)$ variables of the form $m_{t,\star}^{(i)}$ that are set to true. Since $m_{t,\star}^{(i)}$ must take the same value as $m_{t,\star}^{(j)}$, there must be at least $\ell$ variables of the form $m_{t,\star}^{(1)}$ that are set to true. There can be at most one per time step $t$, and so there is exactly one per time step. We consider the TJ-independent set reconfiguration sequence $I_{\sf init}=I_0',\dots,I_\ell'=I_{\sf fin}$  where for $t\in [1,\ell]$ we define $I_t'=I_{t-1}'$ if $m_{t,\emptyset}$ is true, and $I_{t}'=I_{t-1}'\setminus \{v\} \cup \{w\}$ if $m_{t,v,w}^{(1)}$ is true. The subsequence $I_{\sf init}=I_0,\dots,I_T=I_{\sf fin}$ obtained by removing $I_t'$ if $I_t'=I_{t-1}'$, is now a valid TJ-independent set reconfiguration sequence. 

Thus, the result follows from the $W[1]$-completeness of {\sc Weighted 3-CNF-Satisfiability}~\cite{DowneyF99}.
\end{proof}

From the lemma above, the $W[1]$-hardness proof from Mouawad et al.~\cite{MNRSS17}, and the equivalence of the token jumping and token sliding rules
(Theorem~\ref{thm:equivalences}) we can directly conclude the following result.

\begin{corollary}
{\sc Timed TJ-Independent Set Reconfiguration} and {\sc Timed TS-Independent Set Reconfiguration}
with the number of tokens and length of the reconfiguration sequence as combined parameter are $W[1]$-complete.
\end{corollary}

\subsection{Dominating Set}
\label{section:ds-w}
The proof in the previous section can be easily adjusted 
to $\mathrm{W}[2]$-membership for \textsc{Timed TJ-Dominating Set Reconfiguration} with the number of tokens and length of
the reconfiguration sequence as combined parameterization. We remark that the main idea of 
our proof can be applied for several other reconfiguration problems (all that is needed is that
the property of the solution set can be expressed as a CNF formula).

The following problem is complete for $W[2]$ (see e.g. \cite{DowneyF99}).
\begin{verse}
{\sc Weighted CNF-Satisfiability}\\
{\bf Given:} Boolean formula $F$ on $n$ variables in conjunctive normal form; integer $K$.\\
\textbf{Parameter:} $K$.\\
{\bf Question:} Can we satisfy $F$ by setting at most $K$ variables to true?
\end{verse}
We can formulate \textsc{Timed TJ-Dominating Set Reconfiguration} as {\sc Weighted CNF-Satisfiability}, by using the same variables and formulas as we did for \textsc{Timed TJ-Independent Set Reconfiguration}, but changing the last set of formulas that verified whether the solution is an independent set, by the following CNF-formulas that check whether the solution is a dominating set: $\vee_{w\in N[v]} s_{t,w}$ for $t\in [0,\ell]$ and $v\in V$. Here $N[v]$ denotes the set of vertices equal or adjacent to $v$.

Combining the formulation as an instance of {\sc Weighted CNF-Satisfiability}, the NP-hardness result from~\cite{MNRSS17},
and the $W[2]$-completeness of {\sc Weighted CNF-Satisfiability}~\cite{MNRSS17} directly gives the following.

\begin{corollary}
{\sc Timed TJ-Dominating Set Reconfiguration} and {\sc Timed TS-Dominating Set Reconfiguration}
with the number of tokens and length of the reconfiguration sequence as combined parameter are $W[2]$-complete.
\end{corollary}

\section{Conclusion}
\label{section:conclusions}
We showed that for independent set reconfiguration problems parameterized by the number of tokens, the complexity may vary widely depending on the way the length $\ell$ of the sequence is treated. If no bound is given, then we ask for the existence of an undirected path in the reconfiguration graph\footnote{The reconfiguration graph has the possible token configurations as vertex set, and there is an edge between two configurations if we can go from one to the other with a single move.} and indeed the problem is XL-complete. If $\ell$ is given in binary, then we may in particular choose it larger than the maximum number of vertices in the reconfiguration graph, and so this problem is at least as hard as the previous. We show it to be XNL-complete. When $\ell$ is given in unary, it is easier to have a running time polynomial in $\ell$, and indeed the problems becomes XNLP-complete. When $\ell$ is taken as parameter, the problem is $W[1]$-complete. 

On the other hand, switching the rules of how the tokens may move does not affect the parameterized complexity, and the results for dominating set reconfiguration are also similar.
It would be interesting to investigate for which graph classes switching between token jumping and token sliding does affect the parameterized complexities. We give an explicit suggestion below.
\begin{problem}
For which graphs $H$ is \textsc{TJ-Independent Set Reconfiguration} equivalent to \textsc{TS-Independent Set Reconfiguration} under pl-reductions for the class of graphs with no induced $H$?
\end{problem}
The answer might also differ for \textsc{Independent Set Reconfiguration} and \textsc{Dominating Set Reconfiguration}. 
We remark that \textsc{TJ-Clique Reconfiguration} and \textsc{TS-Clique Reconfiguration} have the same complexity for all graph classes \cite{Ito15}.

\subsection*{Acknowledgements}
We would like to thank the referees for useful comments.

\bibliographystyle{plainurl}
\bibliography{xnlp2}

\appendix
\section{Background on XL-, XSL- and XNL-complete Problems}
\label{app:STM}
In this appendix, we discuss why {\sc Accepting Log-Space Symmetric Turing Machine} is XL-complete
(Theorem~\ref{theorem:xsl=xl}).
This equivalence is through a number (well known) steps: by changing the
alphabet, we have a machine with $k$ instead of $k\log n$ cells --- this argument was used
in \cite{ElberfeldST15} in the settings of XL and XNL;
then, we simulate the position of the pointer to the input tape, 
and finally, we apply the classic result
of Reingold~\cite{reingold2008undirected}, yielding XL$=$XSL.


\paragraph*{An XL-complete problem}


Recalling the notation for Nondeterministic Turing Machines from Section~\ref{sec:preliminaries}, we have the following XL-complete problem:

\begin{verse}
    {\sc Input Accepting Binary Log-Space Nondeterministic Turing Machine}\\
    {\bf Given:} An DTM $\mathcal{M} = (\mathcal{S},\Sigma,\mathcal{T},s_{\sf start},\mathcal{A})$ with $\Sigma = \{0,1\}$, a work tape with $k \log n$ cells and input $\alpha \in \Sigma^*$. \\
    {\bf Parameter:} $k$.\\
    {\bf Question:} Does $\mathcal{M}$ accept $\alpha$? 
\end{verse}

\paragraph*{Changing the alphabet.} Note that we can rewrite any Binary Nondeterministic Turing Machine $\mathcal{M}$ with binary alphabet and $k \cdot \log n$ cells on the work tape, to an equivalent NTM $\mathcal{M}'$ with $\Sigma' = [1,n]$ and $k$ cells on the work tape. This can be done by splitting the work tape into $k$ pieces of size $\log n$, viewing each piece as one character $\sigma \in \Sigma$. The transitions can then be re-written accordingly. The equivalence goes both ways: for any cell on the work tape of $\mathcal{M}'$, we can take $\log (n)$ bits to represent its bit encoding on the work tape of $\mathcal{M}$. Whenever a transition changes a cell of $\mathcal{M}'$, a path of $\log n$ transitions is created for $\mathcal{M}$ that changes each bit of the resulting representation accordingly. Hence, the following problem is equivalent (under pl-reductions) to {\sc Input Accepting Binary Log-Space Nondeterministic Turing Machine}.

\begin{verse}
    {\sc Input Accepting Log-Space Deterministic Turing Machine}\\
    {\bf Given:} An DTM $\mathcal{M} = (\mathcal{S},\Sigma,\mathcal{T},s_{\sf start},\mathcal{A})$ with $\Sigma = [1,n]$, a work tape with $k$ cells and input $\alpha \in \Sigma^*$. \\
    {\bf Parameter:} $k$.\\
    {\bf Question:} Does $\mathcal{M}$ accept $\alpha$? 
\end{verse}

\paragraph*{Removing the input tape.} We now argue that this problem is equivalent to {\sc Accepting Log-Space Nondeterministic Turing Machine}. The idea is to track the position of the input tape head by creating extra states. To do this, replace any state $p\in \mathcal{S}$ by states $p^j$ for all $j\in[1,|\alpha|]$. We also replace any transition $\tau \in \mathcal{T}$ by transitions $\tau^j$, for all $j$ such that $\alpha_j$ has the required symbols on the input tape. If, for example, $\tau$ moves the input tape head to the right and transitions from state $p$ to state $q$, then the transition $\tau^j$ goes from state $p^j$ to state $q^{j+1}$. This increases the number of states and transitions with a factor $|\alpha|$. Since $\alpha$ is part of the original input, this increases the size of the NTM by a factor that is polynomial in the size of the input.

\paragraph*{Symmetric Turing Machines.}
The following problem is complete for complexity class XSL:

\begin{verse}
    {\sc Input Accepting Binary Log-Space Symmetric Turing Machine}\\
    {\bf Given:} An STM $\mathcal{M} = (\mathcal{S},\Sigma,\mathcal{T},s_{\sf start},\mathcal{A})$ with $\Sigma = \{0,1\}$, a work tape with $k \log n$ cells and input $\alpha \in \Sigma^*$. \\
    {\bf Parameter:} $k$.\\
    {\bf Question:} Does $\mathcal{M}$ accept $\alpha$? 
\end{verse}

All transformations from previous paragraphs can be  applied to Symmetric Turing Machines 
(and to NTMs, cf.~\cite{ElberfeldST15}). We conclude that {\sc Accepting Log-Space Symmetric Turing Machine} is XSL-complete.\\

\paragraph*{XL $=$ XSL}
Reingold \cite{reingold2008undirected} describes a 
log-space algorithm solving a problem called USTCON. This problem asks, given an undirected graph $G$ and two of its vertices $s$ and $t$, whether $s$ and $t$ are connected. We briefly explain why this result implies L $=$ SL.

We note that L $\subseteq$ SL, because any Deterministic Turing Machine (DTM) can be modelled by a Symmetric Turing Machine (STM) with the same work tape size. For the other direction, one can view the configurations of an STM $\mathcal{M}$ as vertices of a graph $G$ and the allowed transformations between configurations as its edges. Because $\mathcal{M}$ is \emph{symmetric}, these edges always go both ways, so we can view $G$ as an undirected graph. Let $C_{\sf start}$ be the starting configuration and $C_{\sf acc}$ be the accepting configuration of $\mathcal{M}$. 
(We may assume that there exists only one accepting configuration by adding a path of $\mathcal{O}(k\log n)$ states and transition after the accepting state, setting all the symbols on the work tape to 1 and moving the tape head to the first position.)
  Then $\mathcal{M}$ accepts if and only if $C_{\sf start}$ and $C_{\sf acc}$ are connected in $G$. 

We can apply the algorithm of Reingold for USTCON, to compute whether $C_{\sf start}$ and $C_{\sf acc}$ are connected in $G$ in space $\mathcal{O}(\log |V|)$, where $|V|$ is the number of vertices of $G$. The number of different configurations (hence the number of vertices for the corresponding graph) is at most polynomial in $n$, the size of $\mathcal{M}$. (The configurations are specified by the contents of the work space, the position of the head and the current state.)
This implies Reingold's algorithms runs in space $\mathcal{O}(\log n)$, implying that a DTM can decide whether $\mathcal{M}$ accepts in space $\mathcal{O}(\log n)$. 

These ideas also apply to the complexity classes XSL and XL; XL $\subseteq$ XSL because any DTM can be modelled by an STM with the same work tape size. For the other direction, we create a graph $G$ just like before. There are then $n^{k}$ different possibilities for the work tape, $n$ different states and $k$ different tape head positions. Therefore the number of vertices of $G$ is upperbounded by $n^{\mathcal{O}(k)}$. We apply the algorithm of Reingold to compute whether $C_{\sf start}$ and $C_{\sf acc}$ are connected in $G$ in space $\mathcal{O}(k\log n)$. Therefore, a DTM with $\mathcal{O}(k \log n)$ space can compute this, implying XSL $\subseteq$ XL.

Therefore, XL=XSL and any XSL-complete problem is also XL-complete. This concludes the proof of Theorem~\ref{theorem:xsl=xl}.

\end{document}